\documentclass[english]{article}
\usepackage{xcolor}
\usepackage{graphicx}
\usepackage{geometry}
\usepackage[T1]{fontenc}
\usepackage[latin9]{inputenc}
\usepackage{amsbsy}
\usepackage{caption}
\usepackage{subcaption}
\usepackage{setspace}
\usepackage{multirow}
\usepackage{hyperref}
\usepackage[authoryear]{natbib}
\usepackage{amsmath}
\usepackage{amssymb}
\usepackage{booktabs}
\usepackage{bbm}

\newcommand{\RNum}[1]{\uppercase\expandafter{\romannumeral #1\relax}}
\newcommand{\curlybrackets}[1]{\left\{#1\right\}} 
\newcommand{\squarebrackets}[1]{\left[#1 \right]}
\newcommand{\integral}[2]{\int_{#1}^{#2}} 
\newcommand{\parentheses}[1]{\left ( #1\right )}

\begin{document}
\title{Numerical Generalized Randomized Hamiltonian Monte Carlo for piecewise smooth target densities}

\author{Jimmy Huy Tran\footnote{Corresponding author, email: jimmy.tran@uis.no, Department of Mathematics and Physics, University of Stavanger, 4036 Stavanger, Norway} 
  \ and Tore Selland Kleppe\footnote{Department of Mathematics and Physics, University of Stavanger, 4036 Stavanger, Norway}}
\maketitle

\begin{abstract}
Traditional gradient-based sampling methods, like standard Hamiltonian Monte Carlo, require that the desired target distribution is continuous and differentiable. This limits the types of models one can define, although the presented models capture the reality in the observations better. In this project, Generalized Randomized Hamiltonian Monte Carlo (GRHMC) processes for sampling continuous densities with discontinuous gradient and piecewise smooth targets are proposed. The methods combine the advantages of Hamiltonian Monte Carlo methods with the nature of continuous time processes in the form of piecewise deterministic Markov processes to sample from such distributions. It is argued that the techniques lead to GRHMC processes that admit the desired target distribution as the invariant distribution in both scenarios. Simulation experiments verifying this fact and several relevant real-life models are presented, including a new parameterization of the spike and slab prior for regularized linear regression that returns sparse coefficient estimates and a regime switching volatility model. 
\end{abstract}



\doublespacing 
\section{Introduction} \label{sec:intro}
For the majority of modern Bayesian applications, the desired posterior distributions are not analytically available. Numerical and sampling techniques are therefore required to perform posterior inference. Undeniably, the most famous class of such procedures is the Markov chain Monte Carlo (MCMC) methods (see e.g. \citet{GelmanBDA3} for a thorough overview). Lately, extensions in the form of Hamiltonian Monte Carlo (HMC) methods (see e.g \citet{duane1987hybrid, mackenze1989improved} for early contributions) have become popular, mainly as such methods scale much better in higher dimensions \citep[see e.g][]{neal2011mcmc, chen2023does, beskos2013}, leading to more efficient sampling of models with high-dimensional parameter space. For instance, a variant of HMC methods that is used frequently by practitioners is the NUTS algorithm of \citet{hoffman2014no} implemented in the well-known Stan software \citep{JSSv076i01} and PyMC package available in Python \citep{patil2010pymc}. 

Common to conventional MCMC methods are that they are based on reversible Markov processes, which may be less efficient than non-reversible processes \citep{neal2004improving, bierkens2016non}. This motivates the development of MCMC methods based on continuous time piecewise deterministic Markov processes (PDMP) (see e.g. \citet{fearnhead2018piecewise, davis1984piecewise, davis1993markov} for an introduction and a detailed overview). The earlier variants of PDMPs sampler (see \citet{bouchard2018bouncy} for the Bouncy Particle sampler and \citet{bierkens2019zig} for the Zig-Zag sampler) are based on linear deterministic dynamics between events. In order to combine the efficiency of PDMPs and the dimensional scaling of HMC, \citet{bou-rabee_sanz-serna_2018, kleppe2022connecting} examined PDMPs based on deterministic Hamiltonian dynamics. The latter is referred to as the Generalized Randomized Hamiltonian Monte Carlo (GRHMC) process.

The main assumption of all traditional gradient-based methods like the ones mentioned above is that the target distribution must be continuous and differentiable. 
However, for certain models, the ability to capture realistic behavior requires that the posterior distribution either has discontinuous gradients or even is only piecewise continuous. This renders such traditional methods less efficient. Therefore, the present article extends GRHMC to such settings. The first part of this paper concerns the development of a variant of the GRHMC approach that is capable of sampling from continuous target densities with discontinuous gradients at a finite number of boundaries while preserving the order of the numerical integrator. Such scenarios occur in different areas of statistics and machine learning. A relevant example is Bayesian neural networks where the activation function is the rectified linear unit (ReLU) function \citep[see e.g.][]{glorot2011deep, theodoridis2015machine, jospin2022hands}. Another case is a Bayesian regularized linear regression in which a new parameterization that leads to a spike and slab prior with a Dirac delta spike and non-standard slab distribution for the regression parameters is used to obtain similar effects as for standard Lasso regression. As seen later, unlike the Bayesian Lasso \citep{park2008bayesian}, the proposed parametrization of the spike and slab prior leads to samples of the regression parameters that may be identical to zero, and at the same time gives rise to a posterior sampled using the proposed methodology that is continuous, but has discontinuous gradients at given boundaries. For a standard discrete-time HMC algorithm, it is unclear how to resolve this issue as the target is usually required to be smooth everywhere. Recently, \citet{dinh2024hamiltonian} argued that for discrete-time HMC algorithms using e.g. the leapfrog integrator, the method still targets the correct distribution, although much less efficient as the global error of the Hamiltonian is no longer of the classical order $h^2$ \citep{beskos2013}, where $h$ is the step size in the integrator.  However, as shown in this work, such target distributions may be tackled efficiently using a modified GRHMC method in order to take advantage of the dimension scaling in HMC and preserve the global error of the integrator.

Secondly, GRHMC processes targeting piecewise smooth densities are also considered. This setting occurs, for instance, in latent threshold models where the observed quantity switches between states, a common approach in the field of econometrics \citep{nakajima2013bayesian, chang2017new, chang2023oil}. Another example where such a situation arises is presented in \citet{pakman2013auxiliary}. Here, one is interested in sampling from distributions with binary variables using exact HMC. This is done by introducing a set of continuous variables which can be interpreted as a continuous indicator of the binary variable and sampling from this density, which turns out to be piecewise smooth. To resolve the issue with a change in the total energy when a trajectory crosses a boundary of discontinuity, the momentum is altered by either a reflection or refraction process at the point of discontinuity. For more general piecewise smooth targets in the discrete-time HMC setting, \cite{mohasel2015reflection} used the same physical analogy as motivation and considered reflection or refraction when the trajectory obtained by the leapfrog integrator \citep[see e.g.][]{Leimkuhler:2004} hits a boundary between two densities. This is done by modifying the position update step in the leapfrog integrator to detect the boundary crossing event before performing the reflection or refraction operator depending on the momentum and potential energy difference at the intersection point. In contrast, \citet{nishimura2020discontinuous} proposed to switch from Gaussian momentum to Laplace momentum for position coordinates with discontinuities. The numerical integration in this case must therefore also be altered for the reversibility condition to hold, where the coordinates with discontinuities need to be updated coordinate-wise.

Turning to continuous-time PDMPs, \citet{chevallier2021pdmp} provided a general framework and condition for PDMPs to admit a piecewise smooth target density as its invariant distribution. Indeed, the Bouncy Particle sampler by \citet{bouchard2018bouncy} with the transition kernel defined by the reflection and refraction operator whenever the trajectory intersects with a boundary defining discontinuity was shown to satisfy the condition in the same paper. The two situations above naturally translate to the GRHMC process, i.e. GRHMC process inheriting the reflection and refract process as the transition kernel at boundary will also have the target as its invariant distribution. However, a modification to the reflection process is proposed here by introducing a randomized transition kernel similar to \citet{2311.14492}, in which the authors used the kernel for sampling from distributions with restricted domains. It is argued that the modified reflection and refraction kernel also leads to a GRHMC process that admits the correct invariant distribution.

The structure of the paper is as follows: Section \ref{sec:background} introduces the background and setup of GRHMC processes. In Section \ref{sec:discont_grad}, GRHMC processes for sampling from continuous targets with discontinuous gradient are proposed. Similarly, Section \ref{sec:piecewise_smooth} presents GRHMC processes for sampling from piecewise smooth targets. Some simple simulation experiments are performed in Section \ref{sec:simulation}, while Section \ref{sec:reg_lin_reg} applies the proposed parametrization of the spike and slab prior with discontinuous gradient for the linear regression model. Section \ref{sec:bnn} considers a Bayesian neural network with the ReLU activation function and Section \ref{sec:regime_switching} examines a regime switching volatility model using the methodology of Section \ref{sec:piecewise_smooth}. Finally, the paper is concluded with a discussion in Section \ref{sec:discussion}. Source code for the implementations and numerical experiments is available at \url{https://github.com/jihut/grhmc_disc_grad_and_piecewise_smooth}. 

\section{Background and setup} \label{sec:background}

The main goal is to sample from a distribution represented by the probability density function $\pi(\mathbf{q}) \propto c \Tilde{\pi}(\mathbf{q}),\;\mathbf{q} \in \mathbb{R}^d $. Here, $c$ corresponds to an unknown normalizing constant while $\Tilde{\pi}(\mathbf{q})$ is a density kernel of the PDF that allows for evaluation. Unlike the standard sampling situation, $\pi(\mathbf{q})$ considered here will either have discontinuous gradients or be discontinuous itself along a finite number of boundaries. Similar to \citet{kleppe2023, tran2024tuning}, it is often convenient to perform a target standardization by applying the canonical transformation $\mathbf{q} = \mathbf{m} + \mathbf{S} \mathbf{\Bar{\mathbf{q}}}$ \citep[see e.g][]{goldstein2002classical} so that 
\begin{equation} \label{eq:standardized_target_density}
        \Bar{\pi}(\Bar{\mathbf{q}}) = \pi(\mathbf{m} + \mathbf{S}\mathbf{\Bar{\mathbf{q}}})  \lvert \mathbf{S} \rvert \propto \Tilde{\pi}(\mathbf{m} + \mathbf{S}\mathbf{\Bar{\mathbf{q}}}).
\end{equation}
Here, $\mathbf{m}$ can be interpreted as the center vector while $\mathbf{S}$ (which is assumed to be a diagonal matrix with positive entries) represents the scale matrix. In practical applications, these two quantities are usually estimated during the burn-in period. Unless noted otherwise, both $\mathbf{m}$ and $\mathbf{S}$ are adaptively tuned according to the ISG approach in the remaining part of the text \citep[][]{tran2024tuning}.

\subsection{Generalized randomized Hamiltonian Monte Carlo process}
As in any Hamiltonian Monte Carlo methods, $\Bar{\mathbf{q}}$ is recognized as the position vector in a Hamiltonian system together with an auxiliary momentum vector $\Bar{\mathbf{p}}$, which after the canonical transformation (i.e. $\mathbf p=\mathbf S^{-T} \mathbf{\Bar{\mathbf{p}}}$ together with the transformation of $\mathbf{q}$ to $\Bar{\mathbf{q}}$) leads to the following expression of the Hamiltonian total energy: 
\begin{equation} \label{eq:standardized_hamiltonian}
    \mathcal{H}(\mathbf{\Bar{z}}) = \mathcal{H}(\mathbf{\Bar{\mathbf{q}}}, \mathbf{\Bar{\mathbf{p}}}) = - \log{\Bar{\pi}}(\mathbf{\Bar{\mathbf{q}}}) +  \frac{1}{2} \mathbf{\Bar{\mathbf{p}}}^T \mathbf{\Bar{\mathbf{p}}}
\end{equation}
The dynamics of the system is dictated by the Hamilton's equations, which for \eqref{eq:standardized_hamiltonian} are 
\begin{align}
    \mathbf{\Dot{\Bar{q}}}(t) &= \mathbf{\Bar{\mathbf{p}}}(t) \label{eq:standardized_hamiltonian_eq_1}, \\
    \mathbf{\Dot{\Bar{p}}}(t) &= \nabla_{\mathbf{\Bar{\mathbf{q}}}} \log \Bar{\pi}(\Bar{\mathbf{q}}(t)) =  \mathbf{S}\squarebrackets{\nabla_{\mathbf{q}} \log \Tilde{\pi} (\mathbf{q}(t))} \label{eq:standardized_hamiltonian_eq_2}, \quad \mathbf{q}(t) = \mathbf{m} + \mathbf{S} \mathbf{\Bar{q}}(t).
\end{align}

GRHMC processes by \citet{kleppe2022connecting} are PDMPs $\mathbf{\Bar{z}}(t)$, where $\mathbf{\Bar{z}}(t) = \squarebrackets{\mathbf{\Bar{q}}(t)^T, \mathbf{\Bar{p}}(t)^T}^T \in \mathbb{R}^{2d}$, equipped with the Hamiltonian dynamics defined by \eqref{eq:standardized_hamiltonian_eq_1} and \eqref{eq:standardized_hamiltonian_eq_2} as the deterministic dynamics between events. The evolution of $\mathbf{\Bar{z}}(t)$ in time is determined by the flow $\varphi_\tau(\cdot)$ associated with \eqref{eq:standardized_hamiltonian_eq_1} and \eqref{eq:standardized_hamiltonian_eq_2} so that $\mathbf{\Bar{z}}(t + \tau) = \varphi_\tau(\mathbf{\Bar{z}}(t)) $ The event process is allowed to be a non-homogeneous Poisson process with a state-dependent event rate $\lambda(\mathbf{\Bar{z}}) > 0$ as long as certain conditions related to the transition kernel at events are satisfied so that the process admits a stationary distribution with $\Bar{\pi}(\Bar{\mathbf{q}})$ as the marginal distribution of $\Bar{\mathbf{q}}$. For simplicity, in this article, the event rate $\lambda$ is assumed to be independent of the state so that the transition kernel simply refreshes the momentum $\mathbf{\Bar{p}}$ according to $\mathcal{N}(\mathbf{\Bar{p} \mid \mathbf{0}_d, \mathbf{I}_d}$). As with $\mathbf{S}$ and $\mathbf{m}$, this quantity also needs to be tuned to improve the performance. One common measure used for adaptive tuning of $\lambda$ is the "U-turn" time \citep{hoffman2014no}. The setup of the tuning process for this specific purpose of GRHMC is presented further in Appendix \ref{app:tuning_lambda} if this is relevant for a given problem. 

\subsection{Numerical implementation of GRHMC}
Unlike the samplers based on linear deterministic dynamics, e.g. the Bouncy Particle sampler \citep{bouchard2018bouncy}, the Hamiltonian dynamics defined by \eqref{eq:standardized_hamiltonian_eq_1} and \eqref{eq:standardized_hamiltonian_eq_2} usually does not have a closed form solution. Therefore, a numerical ODE solver is required to simulate the dynamics, and hence the corresponding GRHMC process $\mathbf{\Bar{z}}(t)$. This also leads to bias relative to the stationary distribution of the target, but numerical experiments have indicated that the bias can be made small depending on the specified error tolerances in the numerical ODE solver \citep{kleppe2022connecting, kleppe2023}. To obtain $N$ discrete time samples targeting $\pi(\mathbf{q})$ from the simulated process of time length $T$ (after the burn-in period), one can define a uniform sample spacing $\Delta = T / N$ so that $\mathbf{q}_i = \mathbf{m} + \mathbf{S} \mathbf{\Bar{q}(i \Delta)}, \, i = 1, \cdots, N$. 

Similar to \citet{2311.14492}, the ODE solver employed here is the adaptive step size Runge-Kutta solver of order 3(2) by \citet{bogacki19893}. The reason for this choice is essentially the same, mainly motivated by the situations with piecewise smooth densities as this can be seen as a generalization of the restricted domain case. For certain target densities, the numerical GRHMC process could potentially collide with a given boundary frequently, leading to several integration steps being truncated. As lower order solver tends to take smaller time steps, this will prevent larger time steps that require more computational time to be disregarded. 

Finally, the mentioned Runge-Kutta method inherits an embedded root-finding algorithm that can be used to detect momentum refresh events, U-turn events for tuning $\lambda$ presented in  Appendix \ref{app:tuning_lambda} and the boundary crossing events considered later in the text. In summary, Hermite interpolation is used to approximate the state at any given time point during an integration step while preserving the global error of order $h^3$ from the Runge-Kutta solver of order 3 \citep{hairer2008solving}. The interpolation formula can then be used in any root-finding algorithms to find the time point of the potential events, further discussions can be found in \citet{kleppe2023}. 

\section{GRHMC for targets with discontinuous gradients} \label{sec:discont_grad}

\subsection{Setup}
In this section, GRHMC processes that target continuous densities with discontinuous gradients are considered. Assume that the original position space can be divided into $K$ disjoint regions. The different regions are denoted by $R_1, \dots, R_K$, and assumed to be open sets in $\mathbb{R}^d$ so that $\overline{\cup_k R_k} = \mathbb{R}^d$. Each region $R_k$ has its own distinct log-density gradient specified as $\mathbf{g}_k \in \mathbb{R}^d$. In addition, it is convenient to introduce a function $R{(\mathbf{z})}$ that returns an indicator of which region the process is located at a given state, i.e. $R{(\mathbf{z})} = \sum_{k=1}^{K} k \, \mathbbm{1}\squarebrackets{\mathbf{q} \in R_k}$. In most cases considered in this article, it is possible to express $\mathbf{q} \in R_k$ as an inequality of the form $c_k(\mathbf{q}) \geq 0$. The aim is to sample from $\pi$ in which 
\begin{equation} \label{eq:discont_target}
    \nabla_{\mathbf{q}} \log \Tilde{\pi} (\mathbf{q}) = \sum_{k=1}^{K} \mathbf{g}_k \, \mathbbm{1}\squarebrackets{R{(\mathbf{z})}  = k}. 
\end{equation}
In terms of the transformed position coordinates, the region function is given as $R{(\mathbf{\Bar{z}})} = \sum_{k=1}^{K} i \, \mathbbm{1}\squarebrackets{\mathbf{m} + \mathbf{S} \mathbf{\Bar{q}} \in R_k}$ and \eqref{eq:discont_target} becomes 
\begin{equation} \label{eq:standardized_discont_target}
    \nabla_{\mathbf{\Bar{\mathbf{q}}}} \log \Bar{\pi}(\Bar{\mathbf{q}}) =  \mathbf{S}\squarebrackets{\nabla_{\mathbf{q}} \log \Tilde{\pi} (\mathbf{q})} = \mathbf{S} \squarebrackets{\sum_{k=1}^{K} \mathbf{g}_k \, \mathbbm{1}\squarebrackets{R{(\mathbf{\Bar{z}})} = k}}
\end{equation}
with $\Bar{c}_k(\mathbf{\Bar{q}}) = c_k(\mathbf{m} + \mathbf{S} \mathbf{\Bar{q}}) \geq 0$. 
Therefore, each region has its own set of Hamilton's equations \eqref{eq:standardized_hamiltonian_eq_1}-\eqref{eq:standardized_hamiltonian_eq_2} that depends on \eqref{eq:standardized_discont_target}, which induces a specific Hamiltonian flow inside the given region. This flow will be denoted as $\varphi_{T, k}$. 

\subsection{Proposed method for sampling of continuous densities with discontinuous gradients}
Standard Hamiltonian Monte Carlo methods require that there is only one single region, and this region is the whole space itself. Nevertheless, due to the continuous-time approach, one can adopt the following observation from elementary calculus for one-dimensional function: If one is interested in $\integral{a}{b} f(x) \, dx,  \, f \in C^0(\mathbb{R})$, and $f$ is not differentiable at $c \in [a, b]$, the integral can be split into two terms as $\integral{a}{c} f(x) \, dx + \integral{c}{b} f(x) \, dx$. The proposed method for GRHMC processes dealing with scenarios given by \eqref{eq:standardized_discont_target} follows essentially the same principle. The idea is that whenever the numerical ODE solver detects the process crossing a boundary, the remaining part of the integrator step is truncated, and $R{(\mathbf{z})}$ gets updated and alters \eqref{eq:standardized_hamiltonian_eq_2} before the evolution of the system continues with the new gradient until a momentum refresh event (or another boundary crossing). 

In particular, consider a situation where the GRHMC process has evolved for $\tau$ time units so that $\mathbf{\Bar{z}} (\tau)$ defines the current starting point. Denote the overall Hamiltonian flow as $\varphi_{T \mid \mathbf{\Bar{z}}}(\cdot)$ such that $\varphi_{T \mid \mathbf{\Bar{z}}}(\mathbf{\Bar{z}}(\tau)) = \mathbf{\Bar{z}} (\tau + T)$. In addition, the flow induced by \eqref{eq:standardized_hamiltonian_eq_1}-\eqref{eq:standardized_hamiltonian_eq_2} based on the region of the start point is given by $\varphi_{T, R(\mathbf{\Bar{z}}(\tau))}$. This will also be referred to as a region-specific Hamiltonian flow later. Now, when evolving the system by $T$ time units, $\mathbf{\Bar{z}}$ transitions to a different region after $T^* < T$ time units, i.e. $\mathbf{\Bar{z}} (\tau + T^*) = \varphi_{T^*, R(\mathbf{\Bar{z}} (\tau))}(\mathbf{\Bar{z}} (\tau))$ is at a boundary between two regions. Similar to the 1-D integral situation, it is assumed that $\mathbf{\Bar{z}}$ remains unchanged at this time point, but a different flow will be used to evolve the system during the remaining $T - T^*$ time units due to the change in $R(\mathbf{\Bar{z}})$, and hence \eqref{eq:standardized_hamiltonian_eq_1}-\eqref{eq:standardized_hamiltonian_eq_2}, at the boundary point. Representing $\mathbf{\Bar{z}} (\tau + T^*)^-$ and $\mathbf{\Bar{z}} (\tau + T^*)^+$ as the points just before and after the intersection of $\mathbf{\Bar{z}}$ with the boundary, the overall flow is a composition of two region-specific flows assuming that the process only crosses one single boundary during the $T$ time units (note that $R(\mathbf{\Bar{z}} (\tau)) = R(\mathbf{\Bar{z}}(\tau + T^*)^-)$: 

\begin{equation} \label{eq:flow_one_boundary_crossing_event}
    \varphi_{T \mid \mathbf{\Bar{z}}}(\mathbf{\Bar{z}} (\tau)) = \varphi_{T - T^*, R(\mathbf{\Bar{z}}(\tau + T^*)^+)} \circ \varphi_{T^*, R(\mathbf{\Bar{z}}(\tau + T^*)^-)} (\mathbf{\Bar{z}} (\tau))
\end{equation}

Based on the above, one can generalize to a setting of several boundary crossing events occurring in a given time period of the trajectory. This is done by extending the notion of $\varphi_{T \mid \mathbf{\Bar{z}}}$ so that $\varphi_{T \mid \mathbf{\Bar{z}}}$ satisfies 
\begin{equation} \label{eq:flow_many_boundary_crossing_event}
    \varphi_{T \mid \mathbf{\Bar{z}}}(\mathbf{\Bar{z}} (\tau)) = \varphi_{T - t \mid \mathbf{\Bar{z}}} \circ \varphi_{t, R(\mathbf{\Bar{z}}(\tau))} (\mathbf{\Bar{z}} (\tau))
\end{equation}
with $\varphi_{T \mid \mathbf{\Bar{z}}} = \mathbf I_{2d}$ if $T = 0$ and $t = \text{min}(T, T^*)$, where $T^*$ is fully determined by the region at the current initial state $R(\mathbf{\Bar{z}} (\tau))$. 

For a situation with only one single region such that a single flow $\varphi$ is considered, a crucial property of the Hamiltonian dynamics is the time reversibility, which may be expressed as
\begin{equation} \label{eq:involution_single_region_nr1}
    \mathbf R \circ \varphi_{T}  \circ \mathbf R \circ \varphi_{T} = \mathbf I_{2d},
\end{equation}
where $\mathbf R = \text{diag}(\mathbf I_{d}, -\mathbf I_{d})$ is the momentum flip operator. Therefore, when considering a specific region $R_k$, assuming that the region indicator stays constant during the evolution, the following is also true: 
\begin{equation} \label{eq:involution_single_region_nr2}
    \mathbf R \circ \varphi_{T, R(\mathbf{\Bar{z}}_0)}  \circ \mathbf R \circ \varphi_{T, R(\mathbf{\Bar{z}}_0)} = \mathbf I_{2d}
\end{equation}
Here, $\mathbf{\Bar{z}}_0$ denotes the initial state, which in the preceding example is set to be $\mathbf{\Bar{z}}_0 = \mathbf{\Bar{z}}(\tau)$. Time reversibility and the volume preserving property of the flow induced by the Hamiltonian dynamics are some of the crucial attributes to claim that the target distribution is invariant under the evolution of the dynamics. Therefore, in order to argue that the proposed method leads to the correct target distribution, one needs to show that the modified dynamics of switching gradients leads to a flow that is time-reversible and volume preserving. In other words, it is required that the overall Hamiltonian flow also satisfies the condition 
\begin{equation} \label{eq:involution_overall_flow}
    \mathbf R \circ \varphi_{T \mid \mathbf{\Bar{z}}}  \circ \mathbf R \circ \varphi_{T \mid \mathbf{\Bar{z}}} = \mathbf I_{2d}.
\end{equation}
and 
\begin{equation} \label{eq:volume_preserving_overall_flow}
    \left| \text{det}(\nabla_\mathbf{\Bar{z}} \varphi_{T \mid \mathbf{\Bar{z}}} (\mathbf{\Bar{z}})) \right| = 1,
\end{equation}
i.e. the determinant of the Jacobian matrix of the flow with respect to the state variables should have absolute value 1. Appendix \ref{app:time_reverse_overall_flow} argues that the overall Hamiltonian flow indeed inherits both properties. Combined with the results from \citet{kleppe2022connecting}, the proposed modified GRHMC process will admit $\pi$ with discontinuous gradients as its invariant distribution. 

\subsection{Numerical implementation}
\label{sec:numerical_implementation}
It is stated before that the main issue with the traditional HMC method with leapfrog integrator targeting densities with discontinuous gradients is the deviation from the theoretical results of a global error rate of $O(h^2)$ for smooth targets. As mentioned, \citet{dinh2024hamiltonian} showed that the local error occurred after one single leapfrog integrator step is generally of order $h$ based on the jump in the values of the elements of the gradient vector. Thus, the global error of a Hamiltonian trajectory after several leapfrog steps is therefore uncontrollable, which in turn leads to lower acceptance probabilities during the Metropolis step of the HMC algorithm unless the trajectory crosses the boundaries of discontinuous gradients at specific time points \citep{dinh2024hamiltonian}. 

For the GRHMC framework, it is expected that a similar behaviour will occur \emph{if no actions are performed} whenever the trajectory hits the boundaries, i.e. that the global error of a GRHMC trajectory is not of order $h^3$ when using the Runge-Kutta method of order 3(2) by \citet{bogacki19893}. To illustrate these issues and motivate for the proposed method above, which essentially will preserve the order of the global error argued for the setting with smooth targets, consider the two-dimensional distribution given as follows: 
\begin{equation} \label{eq:distribution_for_motivation}
    \begin{split}
        q_1 &\sim N(0,1), \\
        q_2 \mid q_1 &\sim N(\max(0, c q_1), 1).
    \end{split}
\end{equation}
The density of the distribution given above is itself continuous everywhere. However, the gradient is discontinuous at $q_1 = 0$ because $\text{max}(0, x)$ is not differentiable at $x = 0$. Here, $c > 0$ is a constant that can be interpreted as the degree of jump in the gradient, and therefore the complexity of the target due to the discontinuity of the gradient. In this example, the values $c=0.1, 1$ and $10$ are examined. For each value of $c$, three different approaches are used to simulate a trajectory that crosses a boundary of discontinuous gradient only once based on the Hamiltonian dynamics induced by the target density of \eqref{eq:distribution_for_motivation}. When $c = 0.1$ and $1$, the system is evolved for a single time unit, i.e. $T = 1$, with $\mathbf{q}(0) = (-0.5, 1.0)$ and $\mathbf{p}(0) = (1.0, -0.25)$ so that $q_1(1) > 0$ and the trajectory crosses the boundary where the gradient is discontinuous. For $c = 10$, the same initial state is also used, but the system is only evolved for $T = 0.75$ time units so that the trajectory does not hit the boundary more than once. The first method is essentially using the leapfrog integrator from the standard discrete-time HMC method. The second approach is based on the fixed step size part of the Runge-Kutta of order 3(2) algorithm, but where the gradient from both sides of the boundary is used during a single integration step of size $h$. Finally, the third procedure is the proposed method where the same gradient expression is used throughout a single integration step with the boundary collision event located before truncating the remaining part of the step and switching to the gradient given in the new region. 
\begin{figure} 
     \centering
     \includegraphics[width=0.75\textwidth]{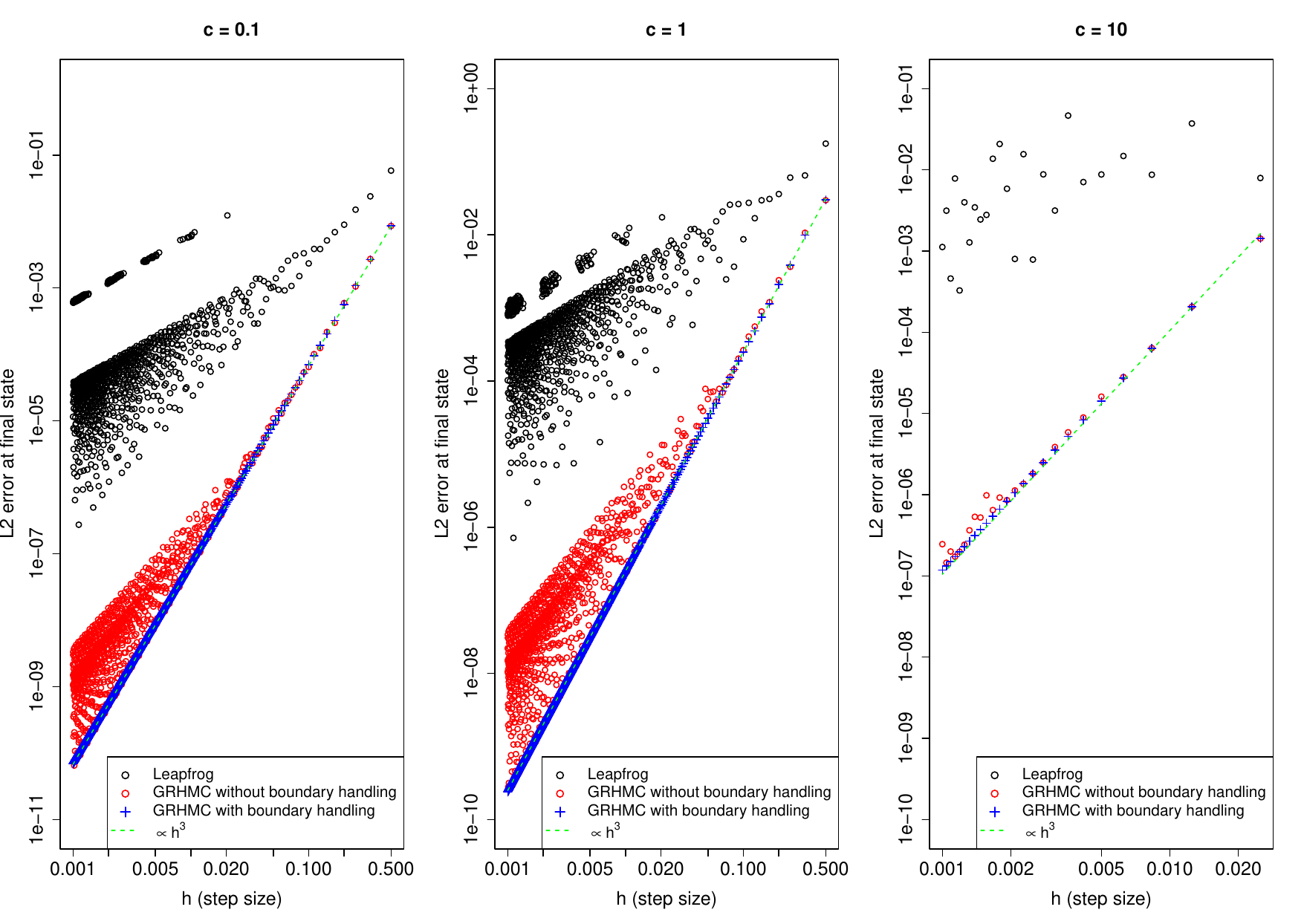}
     \caption{$L_2$-error of the final state $\mathbf{z}(T)=(\mathbf{q}(T), \mathbf{p}(T))$ for three different values of $c$ of the distribution given by \eqref{eq:distribution_for_motivation} plotted against the step size $h$ in a log-log scale. The black dots are the errors obtained by the leapfrog method, the red points and blue crosses represent the errors from the GRHMC procedure without and with proper handling at the boundary of discontinuous gradient.}
     \label{fig:comparison_global_error}
\end{figure}

Figure \ref{fig:comparison_global_error} shows the approximated $L_2$-error of the final state obtained by the different methods with respect to a fixed step size $h$. Here, the approximated true state $\mathbf{z}(t)$ is obtained by solving \eqref{eq:standardized_hamiltonian_eq_1}-\eqref{eq:standardized_hamiltonian_eq_2} using the LSODAR integrator \citep{hindmarsh1982odepack} with very low tolerance for a precise solution. As expected, it is observed that for all three values of $c$, the leapfrog integrator leads to the largest and most varying errors. In contrast, the proposed GRHMC method keeps the error under control while inheriting the $O(h^3)$ global error from the original integrator.  

\begin{figure} 
     \centering
     \includegraphics[width=0.75\textwidth]{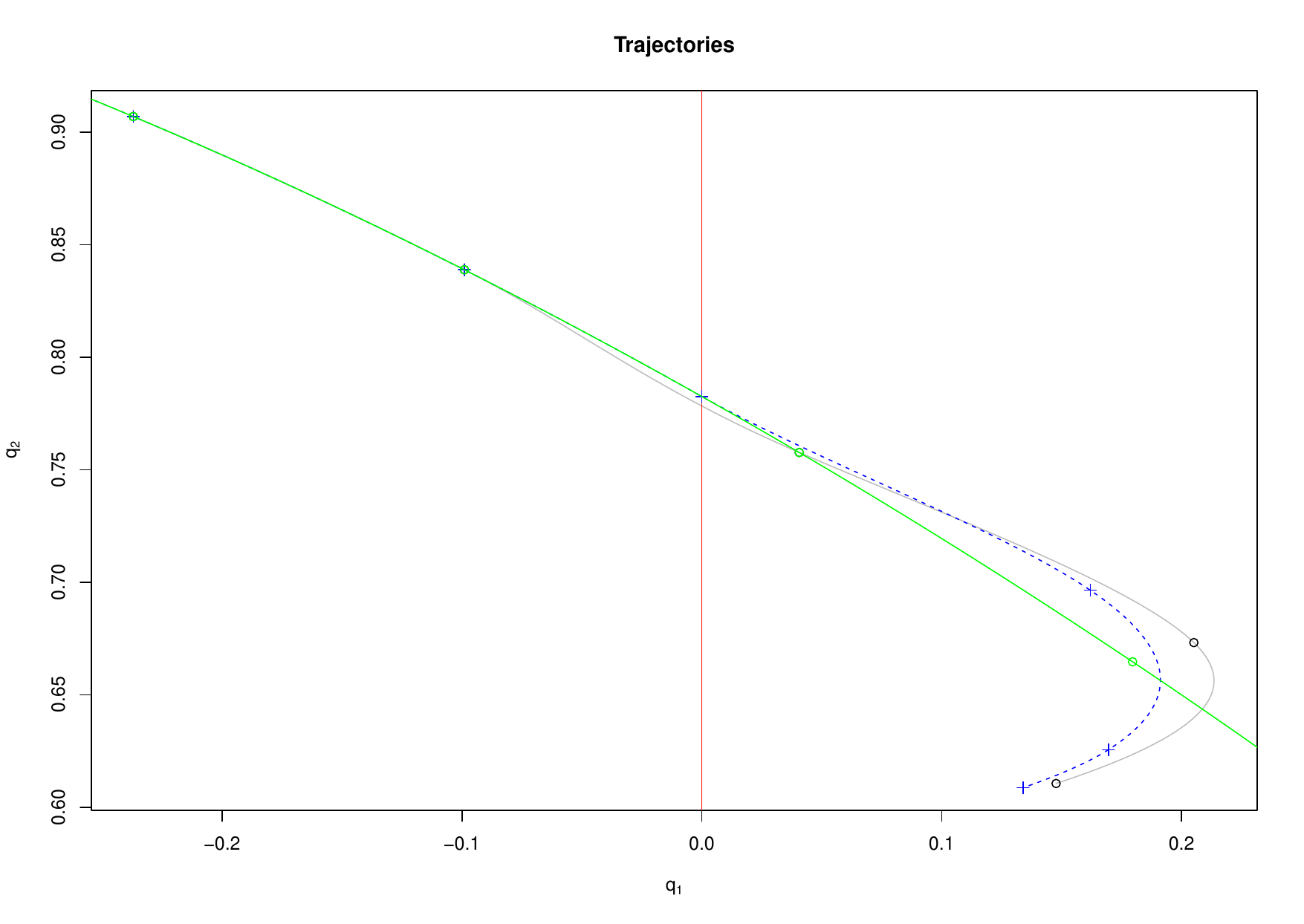}
     \caption{A plot of GRHMC trajectories obtained from different approaches for the example with $c =  10$ and a step size $h = 0.125$. The grey curve indicates the situation where the integrator step is not truncated when colliding with the boundary of discontinuous gradient. The green curve illustrates the continued path of the trajectory if there is no discontinuous gradient boundary and the gradient is equal to the gradient expression from the region $q_1 < 0$. The blue line represents the trajectory obtained from the proposed method. }
     \label{fig:discontinuous_gradient_final_trajectories}
\end{figure}

A more explicit illustration of the issue is presented in Figure \ref{fig:discontinuous_gradient_final_trajectories}. Here, trajectories obtained from different approaches in the scenario with $c = 10$ and step size $h = 0.125$ are plotted. For this example, the grey curve, which corresponds to the situation where the integrator step is not truncated at the boundary, bends more abruptly. In this case, two different gradient expressions are used in one single integration step. On the other hand, the proposed method produces a trajectory with a less sharp turn. Therefore, the final states of the two approaches are slightly different. The $L_2$-error of the final state $\mathbf{z}(0.75)=(\mathbf{q}(0.75), \mathbf{p}(0.75))$ without any modifications is almost four times larger than the error obtained by the proposed method. 

\section{GRHMC for piecewise smooth targets} \label{sec:piecewise_smooth}
\subsection{Setup and background}
Using a similar notation as in Section \ref{sec:discont_grad}, consider a target density of the form 
\begin{equation} \label{eq:piecewise_smooth_density}
    \Bar{\pi}(\mathbf{\Bar{q}}) = \mathbf{S} \sum_{k=1}^K \pi_k(\mathbf{m} + \mathbf{S} \mathbf{\Bar{q}}) \mathbbm{1}\squarebrackets{R{(\mathbf{\Bar{z}})}  = k},
\end{equation}
i.e. each region $R_k$ has its own density $\pi_k$. Although this could be extended to situations where each region density is defined in a different dimension, only the scenario where the dimension across the regions is constant will be of interest in this article. Similar concepts and notations from \citet{chevallier2021pdmp} are also adopted for completeness, see therefore the reference for further details. In that case, denote $p(\mathbf{\Bar{p}})$ as the density of the momentum, which for the standardized GRHMC process is simply $\mathcal{N}(\mathbf{\Bar{p}} \mid \mathbf{0}_d, \mathbf{I}_d)$ \citep{kleppe2022connecting}. As in \citet{chevallier2021pdmp}, defining $\partial R$ as the union of all the boundaries where $\pi$ is discontinuous, one is interested in the situation where $\mathbf{\Bar{q}} \in \partial R$ along two specific regions $R_{k_1(\mathbf{\Bar{q}})}$ and $R_{k_2(\mathbf{\Bar{q}})}$ with a well-defined normal vector (e.g. not vertices that define a rectangle). For simplicity, assume $\pi_{k_1(\mathbf{\Bar{q}})}(\mathbf{\Bar{q}}) < \pi_{k_2(\mathbf{\Bar{q}})}(\mathbf{\Bar{q}})$ with the normal unit vector $\mathbf{\hat{n}}(\mathbf{\Bar{q}}) = \mathbf{n}(\mathbf{\Bar{q}}) / \sqrt{\mathbf{n}(\mathbf{\Bar{q}})^T \mathbf{n}(\mathbf{\Bar{q}})} = (n_1, \cdots, n_d)^T$ pointing towards the region $R_{k_2(\mathbf{\Bar{q}})}$ of higher density and $\mathbf{n}(\mathbf{\Bar{q}}) = \nabla_{\mathbf{\Bar{q}}} \Bar{c}_{k_2(\mathbf{\Bar{q}})}(\mathbf{\Bar{q}}) = \nabla_{\mathbf{{q}}} \curlybrackets{ \mathbf{S}{c}_{k_2(\mathbf{\Bar{q}})}(\mathbf{q})}$. 

Next, define $\mathcal{P}_\mathbf{\Bar{q}}^+$ as the set of momentum vectors that leads to the position vector moving to the higher density region, i.e. $\mathbf{\Bar{p}} \in \mathcal{P}_\mathbf{\Bar{q}}^+$ if $\mathbf{\Bar{p}}^T \mathbf{\hat{n}}(\mathbf{\Bar{q}}) > 0$. Likewise, denote $\mathcal{P}_\mathbf{\Bar{q}}^-$ as the set of momentum vectors that leads to the position vector moving to the lower density region, i.e. $\mathbf{\Bar{p}} \in \mathcal{P}_\mathbf{\Bar{q}}^-$ if $\mathbf{\Bar{p}}^T \mathbf{\hat{n}}(\mathbf{\Bar{q}}) < 0$. In addition, the set of momentum vectors where $\mathbf{\Bar{p}}^T \mathbf{\hat{n}}(\mathbf{\Bar{q}}) = 0$ has in practice Lebesgue measure zero and is therefore not considered in the current setting. Finally, the following density kernel is also defined for the momentum vectors: 
\begin{equation} \label{eq:momentum_density}
    m_\mathbf{\Bar{q}}(\mathbf{\Bar{p}}) = \begin{cases}
        |\mathbf{\Bar{p}}^T \mathbf{\hat{n}}(\mathbf{\Bar{q}})| p(\mathbf{\Bar{p}}) \pi_{k_2(\mathbf{\Bar{q}})}(\mathbf{\Bar{q}}), \text{ if } \mathbf{\Bar{p}} \in  \mathcal{P}_\mathbf{\Bar{q}}^+\\
        |\mathbf{\Bar{p}}^T \mathbf{\hat{n}}(\mathbf{\Bar{q}})| p(\mathbf{\Bar{p}}) \pi_{k_1(\mathbf{\Bar{q}})}(\mathbf{\Bar{q}}), \text{ if } \mathbf{\Bar{p}} \in  \mathcal{P}_\mathbf{x}^-\\
    \end{cases}
\end{equation}

Based on the general framework for any PDMP sampler by \citet{chevallier2021pdmp}, a boundary transition kernel $K_\mathbf{\Bar{q}}(\mathbf{\Bar{p}}' \mid \mathbf{\Bar{p}})$ that only alters the momentum of the state vector yield a PDMP with $\mu(\mathbf{\Bar{z}}) = \Bar{\pi}(\mathbf{\Bar{q}}) p(\mathbf{\Bar{p}}) $ as the invariant distribution if the transition kernel $K'_\mathbf{\Bar{q}}(\mathbf{\Bar{p}}' \mid \mathbf{\Bar{p}}) := K_\mathbf{\Bar{q}}(\mathbf{\Bar{p}}' \mid -\mathbf{\Bar{p}})$ leaves $m_\mathbf{\Bar{q}}(\mathbf{\Bar{p}})$ invariant, i.e. 
\begin{equation} \label{eq:invariant_boundary_kernel}
    m_\mathbf{\Bar{q}}(\mathbf{\Bar{p}}') = \int_{\mathcal{P}_\mathbf{\Bar{q}}^\pm} K'_\mathbf{\Bar{q}}(\mathbf{\Bar{p}}' \mid \mathbf{\Bar{p}}) m_\mathbf{\Bar{q}}(\mathbf{\Bar{p}}) d\mathbf{\Bar{p}}. 
\end{equation}
In short, when denoting $A$ as the generator of the PDMP, $\int Af \, d\mu$ contains an additional term related to the dynamics at the boundary. Since $\int Af \, d\mu = 0$ if $\mu$ is an invariant distribution \citep[see e.g.][]{davis1984piecewise}, \citet{chevallier2021pdmp} showed that the condition \eqref{eq:invariant_boundary_kernel} must be satisfied in order for this boundary term to be equal to zero. 

\subsection{Modified reflection and refraction kernel}
It is mentioned before that the reflection and refraction process provided by \citet{mohasel2015reflection} is valid for the discrete-time HMC situation after a small modification of the leapfrog integrator. Transitioning to the continuous-time scenario, \citet{chevallier2021pdmp} showed that the Bouncy Particle sampler inheriting the same deterministic kernel also satisfies \eqref{eq:invariant_boundary_kernel}. More specifically, if $\mathbf{\Bar{q}} \in \partial R$, let $\Delta U = \log(\pi_{k_2(\mathbf{\Bar{q}})}(\mathbf{\Bar{q}})) - \log(\pi_{k_1(\mathbf{\Bar{q}})}(\mathbf{\Bar{q}})) > 0$ be the difference in the potential energy between the lower and higher density region in consideration. The boundary transition kernel $K_\mathbf{\Bar{q}}(\mathbf{\Bar{p}}' \mid \mathbf{\Bar{p}})$ based on reflection and refraction is given by: 
\begin{equation} \label{eq:reflection_refraction_process}
    \mathbf{\Bar{p}}' = \begin{cases}
        \mathbf{\Bar{p}} + \parentheses{ \sqrt{(\mathbf{\Bar{p}}^T \mathbf{\hat{n}}(\mathbf{\Bar{q}}))^2 + 2 \Delta U} - \mathbf{\Bar{p}}^T \mathbf{\hat{n}}(\mathbf{\Bar{q}})} \mathbf{\hat{n}}(\mathbf{\Bar{q}}), \text{ if } \mathbf{\Bar{p}} \in  \mathcal{P}_\mathbf{\Bar{q}}^+ \\
        \mathbf{\Bar{p}} + \parentheses{- \sqrt{(\mathbf{\Bar{p}}^T \mathbf{\hat{n}}(\mathbf{\Bar{q}}))^2 - 2 \Delta U} - \mathbf{\Bar{p}}^T \mathbf{\hat{n}}(\mathbf{\Bar{q}})} \mathbf{\hat{n}}(\mathbf{\Bar{q}}), \text{ if } \mathbf{\Bar{p}} \in  \mathcal{P}_\mathbf{\Bar{q}}^- \text{ and } (\mathbf{\Bar{p}}^T \mathbf{\hat{n}}(\mathbf{\Bar{q}}))^2 > 2 \Delta U\\
        \mathbf{\Bar{p}} - 2 (\mathbf{\Bar{p}}^T \mathbf{\hat{n}}(\mathbf{\Bar{q}})) \mathbf{\hat{n}}(\mathbf{\Bar{q}}), \text{ if } \mathbf{\Bar{p}} \in  \mathcal{P}_\mathbf{\Bar{q}}^- \text{ and } (\mathbf{\Bar{p}}^T \mathbf{\hat{n}}(\mathbf{\Bar{q}}))^2 < 2 \Delta U\\
    \end{cases}
\end{equation}
The first situation corresponds to the refraction procedure when the position vector moves from the lower density region $k_1(\mathbf{\Bar{q}})$ to $k_2(\mathbf{\Bar{q}})$, thus increasing the velocity along the direction of the normal vector. The second scenario is a refraction when the momentum component parallel to the normal vector is large enough to cross the potential boundary from the higher to the lower density region, thus reducing the momentum along the component of the momentum vector that is parallel to the normal vector. In this case, the component points in the opposite direction of $\mathbf{\hat{n}}(\mathbf{\Bar{q}})$ and therefore leads to the minus sign in front of the refraction factor. Finally, if the momentum component parallel to the normal vector is not large enough to cross the potential boundary from the higher to the lower density region, a reflection occurs so that the momentum points back towards the higher density region.

Although the result that the kernel above obeys \eqref{eq:invariant_boundary_kernel} for the Boundary Particle sampler, the proof presented by \cite{chevallier2021pdmp} only assumed that  $p(\mathbf{\Bar{p}})$ follows a normal distribution. Thus, the result is immediately valid for the GRHMC process, which is naturally intuitive since the kernel was originally proposed for the discrete-time HMC method. Nevertheless, assume now that the momentum refresh rate $\lambda$ is tuned using the procedure based on the NUTS condition given in Appendix \ref{app:tuning_lambda}. In certain scenarios where boundary collisions occur frequently, the resulting adaptive $\lambda$ after the burn-in period could still be slightly too small such that the deterministic kernel might lead to trajectories covering the same path several times. Setting a predefined value of $\lambda$ requires manual effort and could lead to a large choice of $\lambda$, which is also not very efficient as the trajectory will exhibit some random walk behavior \citep{kleppe2022connecting}. One way of adding more randomness without increasing $\lambda$ is to use a randomized reflection kernel. More specifically, the refraction parts of \eqref{eq:reflection_refraction_process} are the same while the reflection process is now given by 
\begin{equation} \label{eq:randomized_reflection}
    \mathbf{\Bar{p}}' = \mathbf{x} - \parentheses{(\mathbf{\Bar{p}} + \mathbf{x})^T \mathbf{\hat{n}}(\mathbf{\Bar{q}})} \mathbf{\hat{n}}(\mathbf{\Bar{q}}), \, \mathbf{x} \sim N(\mathbf{0}_d, \mathbf{I}_d). 
\end{equation}
This is the same kernel based on the $(d-1)$-dimensional standard normal distribution on the space orthogonal to the normal vector presented by \citet{2311.14492} \citep[see also][]{wu2017generalized, bierkens2018piecewise,bierkens2023methods}, who used this kernel for targets with restricted domains. The kernel will still reflect the momentum component tangential to the normal vector, but it allows for a change in the component parallel to the boundary. It is shown in Appendix \ref{app:show_randomized_reflection_piecewise} that the proposed modification to the original reflection will satisfy \eqref{eq:invariant_boundary_kernel}. Therefore, the sparse version of the randomized reflection kernel, where only the dimensions with non-zero normal components at a given boundary will be refreshed using \eqref{eq:randomized_reflection}, is also valid to attain the correct invariant distribution. 

\begin{figure} 
     \centering
     \includegraphics[width=0.9\textwidth]{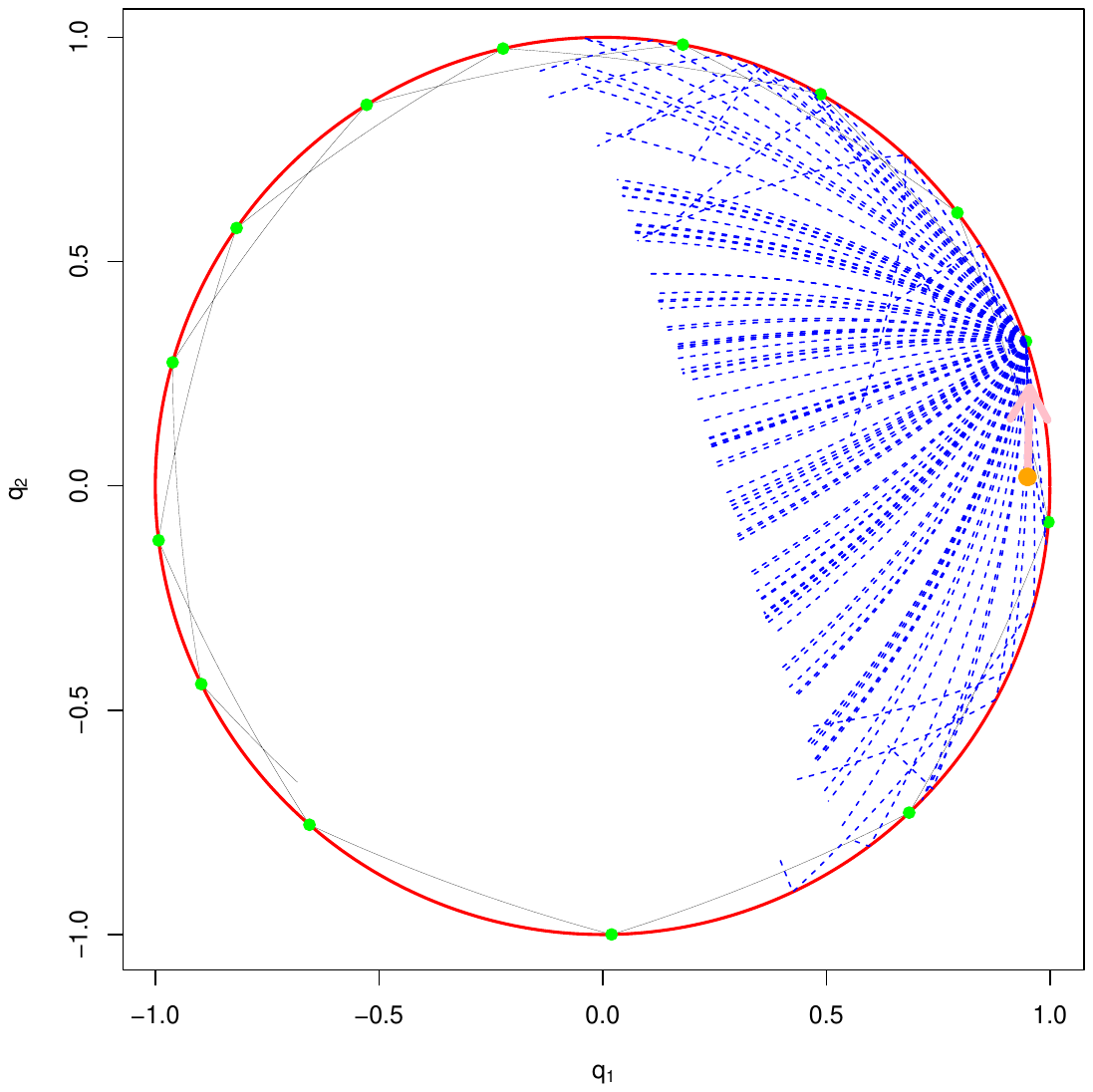}
     \caption{An illustration of a situation of repeated boundary collisions for the target given by \eqref{eq:bivariate_normal_unit_circle_boundary}. The dark line corresponds to the trajectory obtained by the deterministic reflection kernel, evolving the system for $T = 5$ time units. The green dots are the points of collisions of this trajectory. The blue striped lines are initial parts of 100 different trajectories due to the randomized kernel. The red curve is the boundary itself, with the orange point being the initial values of $\mathbf{q}$ and the pink arrow representing the initial momentum. }
     \label{fig:bivariate_normal_unit_circle_boundary_frequent_collision}
\end{figure}

The following illustration shows a scenario where using the randomized kernel might be beneficial. Consider a target distribution given by \eqref{eq:bivariate_normal_unit_circle_boundary}. Assume a situation where the momentum refresh event rate is set to a small value. For simplicity, assume that after a momentum refresh event, $\mathbf{q} = c(0.95, 0.02)^T$ and $\mathbf{p} = c(0.05, 2)^T$. Then, unless the momentum vector is sampled again in a short period of time, repeated boundary collisions will occur as presented in Figure \ref{fig:bivariate_normal_unit_circle_boundary_frequent_collision}. The resulting trajectory from the deterministic reflection kernel will always lie in a certain neighbourhood close to the boundary. On the other hand, the randomized kernel does not inherit this issue. 

\section{Simulation experiments} \label{sec:simulation}
In this section, simulation experiments are presented with the purpose of illustrating the proposed methods for the two distinct settings described earlier. 
\subsection{A continuous target with discontinuous gradient}
First, a continuous target density with discontinuous gradient is examined. Consider the following two-dimensional distribution represented by 
\begin{equation} \label{eq:max_model}
    \begin{split}
        q_1 &\sim N(0,1), \\
        q_2 \mid q_1 &\sim N(\max(0, q_1), 1), 
    \end{split}
\end{equation}
i.e. the target is simply \eqref{eq:distribution_for_motivation} with $c = 1$. 
The marginal probability density function (PDF) of $q_2$ is given by 
\begin{equation} \label{eq:max_model_true_pdf}
    f_2(q_2)=\frac{1}{2}\parentheses{\phi(q_2) + \sqrt{2} \Phi(\sqrt{2} q_2 / 2)  \phi(\sqrt{2} q_2 / 2)},
\end{equation}
where $\phi$ and $\Phi$ denote the probability density function (PDF) and cumulative density function (CDF) of the standard normal distribution, respectively. 
The joint PDF of $q_1$ and $q_2$ is continuous everywhere, but it has discontinuous gradients along the line $q_1 = 0$. More specifically, 
\begin{equation}
    \nabla_{\mathbf{q}} \log {\pi} (\mathbf{q}) = \begin{cases}
        (-q_1, -q_2)^T, \, q_1 < 0, \\
        (q_2 - 2q_1, q_1-q_2)^T, q_1 > 0.
    \end{cases}
\end{equation}
Therefore, in order to sample from $\pi$, the GRHMC process described in Section \ref{sec:discont_grad} could be applied. For this purpose, 10 independent trajectories with the sampling time duration $T = 100000$ and $N = 50000$ are simulated for each scenario. Before the sampling period, the trajectories are also run for 10000 time units, where $\mathbf{m}$, $\mathbf{S}$ and $\lambda$ are adaptively tuned. Here, the absolute and relative tolerances of the Runge-Kutta 3(2) integrator are chosen to be $tol_a=tol_r=10^{-4}$. The results are obtained by merging all the samples from the 10 trajectories. The final number of samples will therefore be $500000$. 
\begin{figure} 
     \centering
     \includegraphics[width=0.75\textwidth]{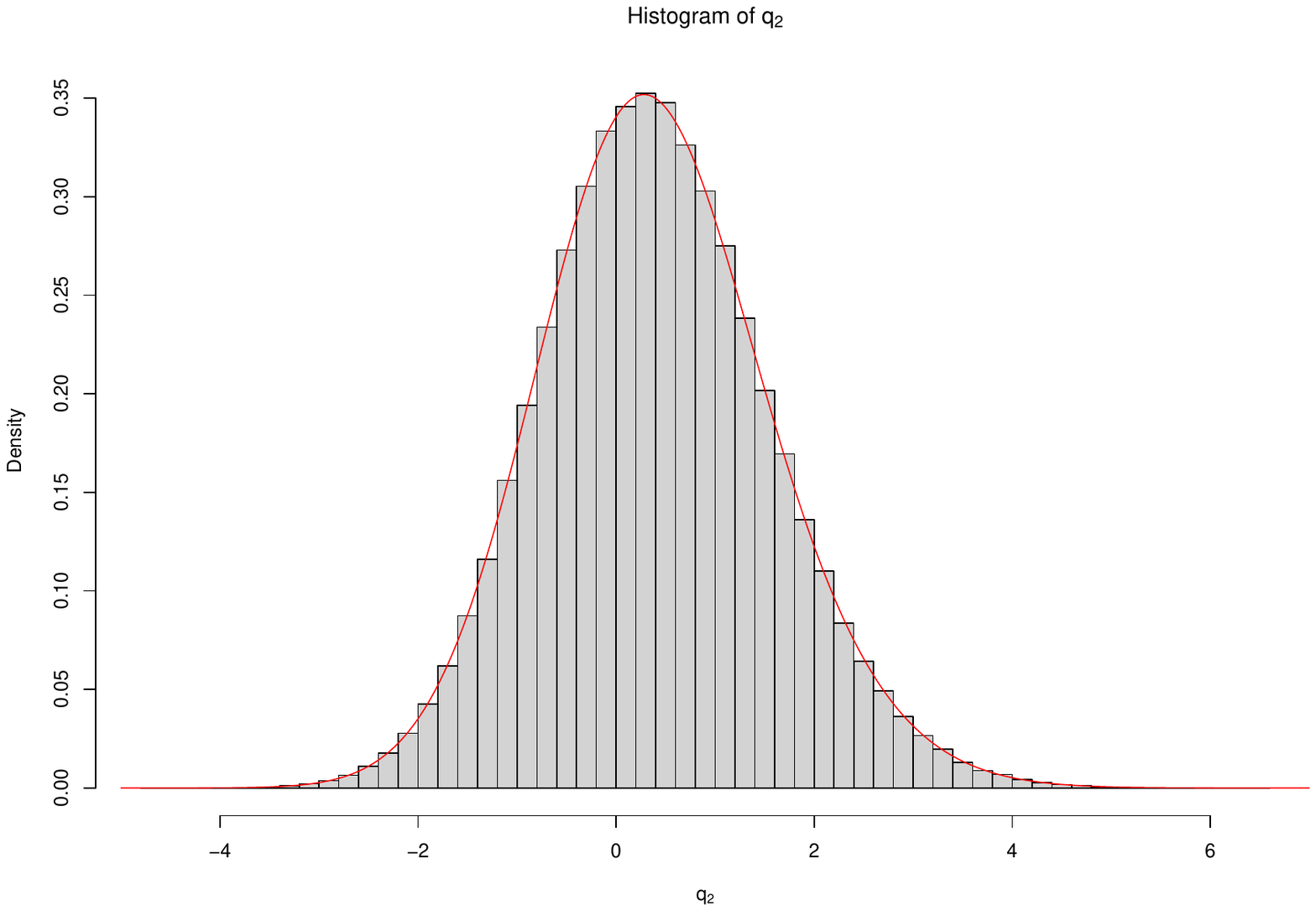}
     \caption{Histogram corresponding to the 500000 samples obtained from the 10 independent trajectories targeting the distribution characterized by \eqref{eq:max_model}. The red curve corresponds to the true PDF \eqref{eq:max_model_true_pdf}.}
     \label{fig:max_model_q2_pdf}
\end{figure}

From Figure \ref{fig:max_model_q2_pdf}, it is observed that the histogram based on the samples obtained from the 10 independent trajectories simulated using the proposed methodology matches very well to the true PDF. Thus, this indicates that the GRHMC process does indeed admit the target defined by \eqref{eq:max_model} as its stationary distribution, modulus imperceivable numerical errors.

\subsection{A piecewise smooth target}
In this subsection, a piecewise smooth target density is investigated. For this example, consider the following two-dimensional target density
\begin{equation} \label{eq:bivariate_normal_unit_circle_boundary}
    \pi(\mathbf{q}) = 
    \begin{cases}
        \mathcal{N}(\mathbf{q} \mid \mathbf{0}_2, \mathbf{I}_2), \, \lvert \mathbf{q} \rvert < 1, \\
        \frac{c_1}{c_2}\mathcal{N}(\mathbf{q} \mid \mathbf{0}_2, 4 \mathbf{I}_2), \, \lvert \mathbf{q} \rvert > 1,
    \end{cases}
\end{equation}
where $c_1 = \int_{\lvert \mathbf{q} \rvert > 1} \mathcal{N}(\mathbf{q} \mid \mathbf{0}_2, \mathbf{I}_2) \, d\mathbf{q} = \exp{(-1/2)}$ and $c_2 = \int_{\lvert \mathbf{q} \rvert > 1} \mathcal{N}(\mathbf{q} \mid \mathbf{0}_2, 4 \mathbf{I}_2) \, d\mathbf{q} = \exp{(-1/8)}$.

\begin{figure} 
     \centering
     \includegraphics[width=0.75\textwidth]{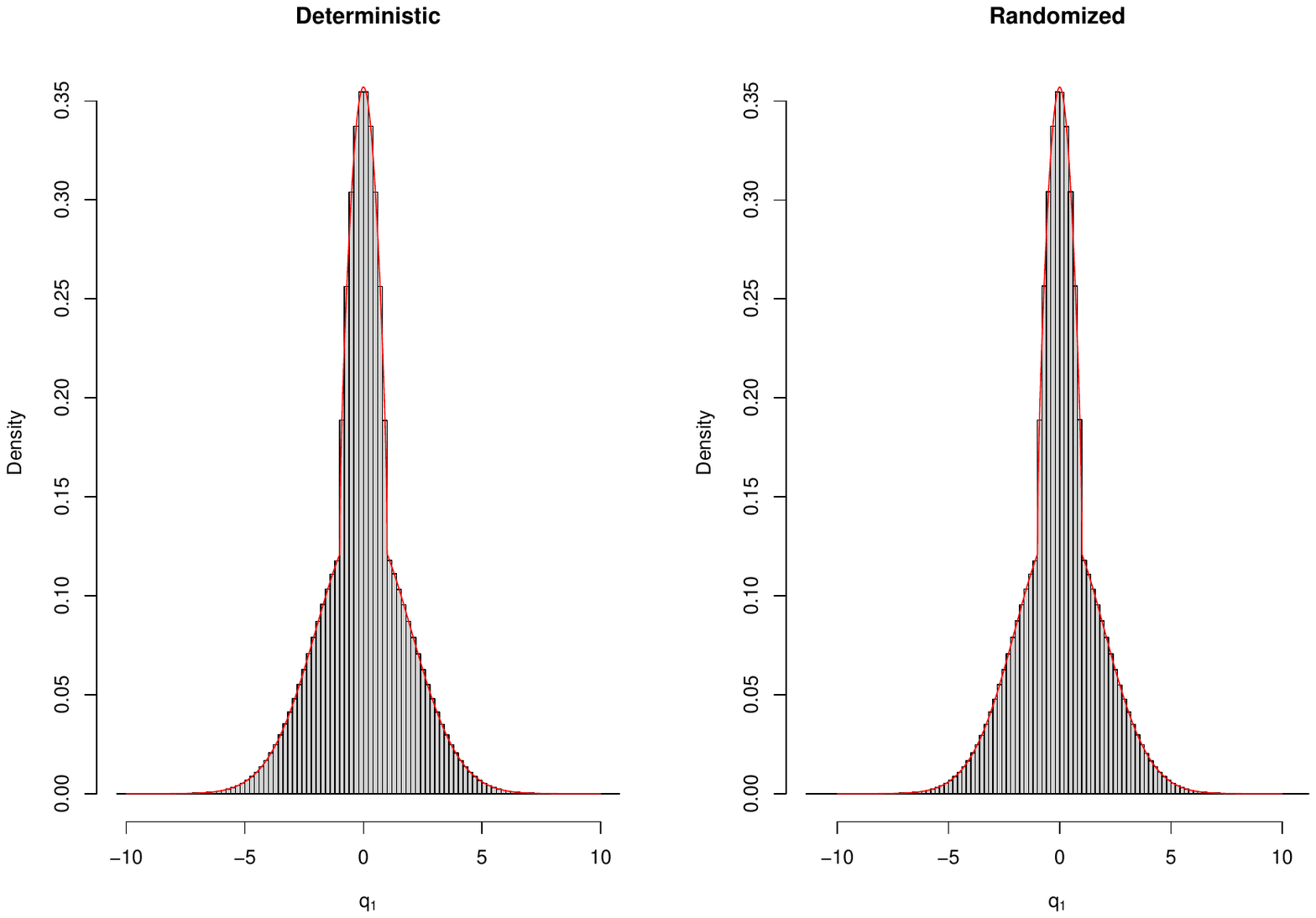}
     \caption{Histograms of the samples obtained by aggregating 1000 independent trajectories targeting \eqref{eq:bivariate_normal_unit_circle_boundary}. Left plot: Marginal samples of $q_1$ obtained using deterministic reflection kernel. Right plot: Marginal samples of $q_1$ obtained using randomized reflection kernel. The red curve corresponds to the true marginal density of $q_1$ given by \eqref{eq:bivariate_normal_unit_circle_boundary_marginal}. }
     \label{fig:histogram_bivariate_normal_unit_circle_boundary_deterministic_vs_randomized}
\end{figure}

It can be shown that the marginal density of $q_1$ (and also of $q_2$ due to spherical symmetry) is 
\begin{equation} \label{eq:bivariate_normal_unit_circle_boundary_marginal}
    \pi_1(q_1) = 
    \begin{cases}
        2\frac{c1}{c2} \parentheses{1 - \Phi\parentheses{\frac{\sqrt{1-q_1^2}}{2}}} \mathcal{N}(q_1 \mid 0, 2^2) + \mathcal{N}(q_1 \mid 0, 1) \parentheses{2 \Phi\parentheses{\sqrt{1-q_1^2}} - 1} , \, \lvert q_1 \rvert < 1, \\
        \frac{c_1}{c_2} \mathcal{N}(q_1 \mid 0, 2^2), \, \lvert q_1 \rvert > 1.
    \end{cases}
\end{equation}
To verify that both the deterministic and the randomized reflection kernel lead to a GRHMC process that produces samples from the correct target distribution, 1000 independent trajectories targeting \eqref{eq:bivariate_normal_unit_circle_boundary} are simulated for each of the two reflection methods with $T = 100000$, $N = 100000$ and $\lambda = 0.2$ and the marginal samples of $q_1$ are compared against \eqref{eq:bivariate_normal_unit_circle_boundary_marginal}. Also, $\mathbf{m}$ and $\mathbf{S}$ are set to the zero vector and identity matrix. Figure \ref{fig:histogram_bivariate_normal_unit_circle_boundary_deterministic_vs_randomized} indicates that both kernels lead to samples from the correct marginal distribution of $q_1$. 

\section{Numerical illustrations} \label{sec:numerical_illustrations}

\subsection{Regularized linear regression} \label{sec:reg_lin_reg}
In this section, a new parameterization of the spike and slab prior with a Dirac delta spike that leads to a continuous target with discontinuous gradients is illustrated for the setting of regularized linear regression. It is well known that estimates obtained from traditional regularization methods like the Lasso and ridge regression can be interpret as the maximum a posteriori (MAP) of a Bayesian linear regression with specific choices of the prior distributions on the coefficients \citep{james2013introduction}. However, in the classical sense of sampling from the posterior, none of the above yields samples of the coefficients that are identical to zero. Several proposals of different continuous shrinkage priors have appeared since then, for instance, the Bayesian Lasso \citep{park2008bayesian} and horseshoe prior \citep{carvalho2009handling}, but common to these is that MCMC samples cannot be exactly zero and explicit variable selection is therefore not performed. On the other hand, under the original parameterization of the spike and slab prior with a Dirac delta spike \citep{lempers1971posterior, mitchell1988bayesian,george1993variable,ishwaran2005spike}, variable selection is enabled by the introduction of a discrete latent variable that indicates if a regression coefficient is zero or not. However, the discrete nature of this parameter leads to complications for gradient-based sampling methods that require an unconstrained continuous parameter. 

Given the response vector $\mathbf{Y} \in \mathbb{R}^n$ and the covariate matrix $\mathbf{X} \in \mathbb{R}^{n \times p}$ (both are assumed to be centered and scaled in this case so that the intercept can also be omitted, transforming back to the original scale where the intercept is relevant can be done later using the observations from Appendix \ref{app:details_reg_lin_reg}), consider now the standard linear regression setup: 
\begin{equation}
    \mathbf Y \sim N(\mathbf X \boldsymbol{\beta},\sigma^2 \mathbf I) \label{eq:standard_lin_reg_eq}
\end{equation}
Here, $\boldsymbol{\beta}^T = (\beta_1, \dots, \beta_p)$ is the regression coefficient vector. It is of interest to perform variable selection in a Bayesian approach by obtaining a large ratio of posterior samples equal to zero for the corresponding coefficient of a variable if the variable itself is not strongly related to the response. To this end, a prior for $\beta_i$ is defined via: 
\begin{gather}
    \beta_i = \max(0,\beta_i^+) - \max(0,\beta_i^-), \; i=1, \dots, p, \label{eq:beta_wrt_beta_plus_and_minus} \\
    \beta^+_i \sim N(\mu,\rho^2), \label{eq:beta_plus_prior} \\
    \beta^-_i \sim N(\mu,\rho^2) \label{eq:beta_minus_prior}
\end{gather}
With all $\beta_i^+$ and $\beta_i^-$ being independent, the prior hyperparameters $\mu$ and $\rho$ combined can be thought of as the global degree of shrinkage. The model is completed with a prior of $\sigma$, which after reparameterization to an unconstrained space through $\gamma = \log(\sigma^2)$ is chosen to be $\gamma \sim N(0,1)$ for illustration purposes. Finally, by introducing $(\boldsymbol{\beta}^+)^T = (\beta_1^+, \dots, \beta_p^+)$ and $(\boldsymbol{\beta}^-)^T = (\beta_1^-, \dots, \beta_p^-)$, the main objective is to sample $\boldsymbol{\beta}^+, \boldsymbol{\beta}^-$ and $\gamma$ based on the presented setup, which are all unconstrained parameters defined everywhere on the real line so that Hamiltonian Monte Carlo methods can be applied. However, even if the posterior is everywhere continuous with respect to all parameters, it also admits now discontinuous gradients with respect to $\boldsymbol{\beta}^+$ and $ \boldsymbol{\beta}^-$ due to  the ReLU function in \eqref{eq:beta_wrt_beta_plus_and_minus} for the relation between the actual regression coefficient $\beta_i$ and $\beta_i^+$ (similarly for $ \beta_i^-$). 
Although standard HMC methods are in theory not suitable for this setup, the proposed GRHMC method from Section \ref{sec:discont_grad} is intended for sampling from targets like $\pi(\boldsymbol{\beta}^+, \boldsymbol{\beta}^-, \gamma \mid \mathbf{Y}, \mathbf{X}, \mu, \rho)$ . 

From \eqref{eq:beta_wrt_beta_plus_and_minus}, one can observe that the regression coefficient $\beta_i$ is identical to zero whenever $\beta_i^+ < 0$ and $\beta_i^- < 0$. More specifically, the choice of $\mu$ and $\rho$ determines the prior $P(\beta_i = 0$), $\text{Var}(\beta_i)$ and $\text{Var}(\beta_i \mid \beta_i \neq 0)$, which in turn impacts the shrinkage effect. The details of deciding $\mu$ and $\rho$ based on prespecified values of prior $P(\beta_i = 0$) and $\text{Var}(\beta_i \mid \beta_i \neq 0)$ are presented in Appendix \ref{app:details_reg_lin_reg}. As an example, if $\mu = 0$ and $\rho = 1$ (or any $\rho > 0$), the prior $P(\beta_i = 0$) is simply $P(\beta_i^+ < 0)P(\beta_i^- < 0)= 0.5 ^ 2 = 0.25$. Therefore, the proposed model implies variable selection while still sampling from continuous variables. In fact, based on $\beta^+_i$ and $\beta^-_i$, one can represent the prior of $\beta_i$ as 
\begin{equation}
    p(\beta_i \mid \chi_i) = \chi_i\delta_0(\beta_i) + (1-\chi_i) p_{\chi_i=1}(\beta_i),
\end{equation}
where $\chi_i \in \curlybrackets{0, 1}$, $P(\chi_i=1)=\Phi(-\mu/\rho)^2$, $\delta_0(\cdot)$ is the Dirac delta function centred at zero and $p_{\chi_i=1}(\cdot)$ is a non-standard slab distribution that is omitted in this article for future works. 

\begin{figure} 
     \centering
     \includegraphics[width=0.9\textwidth]{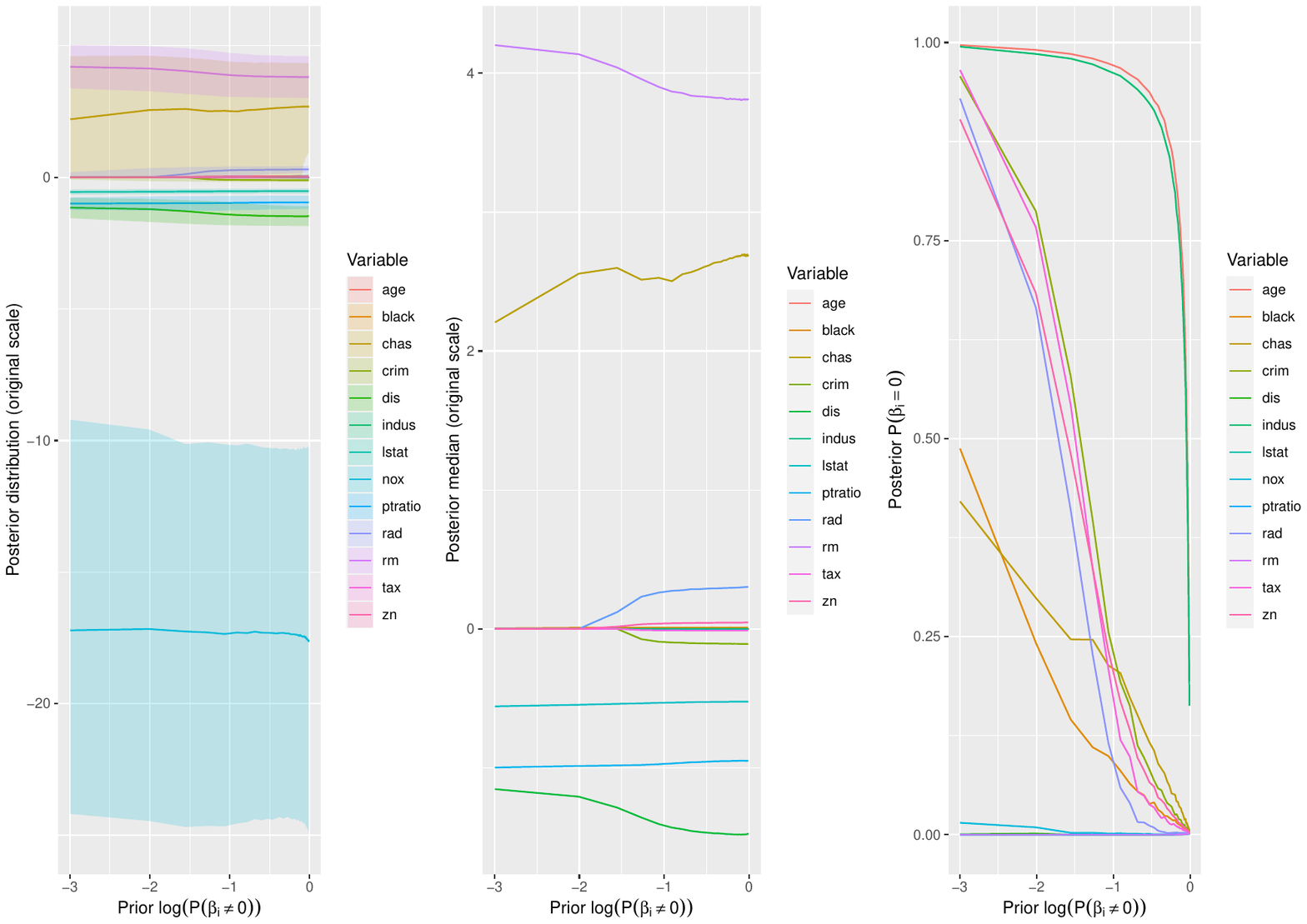}
     \caption{Plots of the model given by \eqref{eq:standard_lin_reg_eq}-\eqref{eq:beta_minus_prior} fitted to the Boston housing dataset with respect to the prior $P(\beta_i = 0)$ when $\text{Var}(\beta_i \mid \beta_i \neq 0) = 1$. The left plot displays the posterior median with the 95\% credible interval of the regression coefficients in the original scale. The middle plot shows the coefficients with smaller scale to examine the shrinkage effect when varying $P(\beta_i = 0)$. Finally, the right plot represents the ratio of posterior samples of the coefficients that are equal to zero, with a similar interpretation as the posterior inclusion probability from Bayesian variable selection.}
     \label{fig:lin_reg_boston}
\end{figure}

As an explicit example, the model is fitted on the classical Boston housing dataset \citep[see e.g. ][]{james2013introduction}, which contains $n = 506$ observations and $p = 13$ covariates. For this illustration, the prior conditional variance $\text{Var}(\beta_i \mid \beta_i \neq 0)$ is fixed to 1 while the prior $P(\beta_i = 0$) varies between 0.01 and 0.95 as the degree of shrinkage. The values of $\mu$ and $\rho$ such that $\text{Var}(\beta_i \mid \beta_i \neq 0) = 1$ and $P(\beta_i = 0) = \alpha$ for a given choice of $\alpha$ between 0.01 and 0.95 are obtained by solving the two conditions numerically. Figure \ref{fig:lin_reg_boston} shows the result of aggregating 10 independent trajectories with $T=10000$ and $N=10000$ for a given value of $P(\beta_i = 0$) over the specified range without any adaptive procedure or warm-up period (for illustration purposes). Here, the momentum refresh rate of the GRHMC process is therefore set to $0.2$ while $\mathbf{m} = \mathbf{0}_{2p+1}$ and $\mathbf{S} = \mathbf{I}_{2p+1}$. 

From the first two plots, it is observed that increasing the value of prior $P(\beta_i = 0$) leads to posterior medians approaching zero for a certain set of coefficients/variables. The same can be seen from the right most plot, with the posterior probability of $\beta_i$ equal to zero growing when the prior $P(\beta_i = 0$) increases. Note, however, that there are certain variables that tend to have a very low posterior $P(\beta_i = 0$) even with a large prior $P(\beta_i = 0$). In order to shrink and achieve a high posterior probability of zero for every variable, one needs to set a very high $P(\beta_i = 0$) with small $\text{Var}(\beta_i \mid \beta_i \neq 0)$. Nevertheless, this implies that the dynamics is restricted to a very small region outside of the prior, thus leading to posterior samples that behave similar to the priors. This is an indication that this particular set of covariates might have a strong explanatory power on the response.

\subsection{Bayesian Neural Network with ReLU activation function} \label{sec:bnn}
Another example of a situation where the target density is continuous with discontinuous gradients is the Bayesian neural network with the ReLU function as the activation function. Similar to the regression model discussed earlier, the discontinuity is caused by the ReLU function itself. 

Given the response variable $Y_j$ and a covariate vector $\mathbf{X}_j = (X_{j1}, \dots X_{jp})^T$,  the following simple model with only one hidden layer is considered: 
\begin{gather}
    Y_j \sim N(\mu_j, \sigma^2), \\
    \mu_j = \alpha + \sum_{k = 1}^K w_k g(\eta_{kj}), \\
    \eta_{kj} = \delta_k + \mathbf{X}_j^T \boldsymbol{\beta}_k, \, \boldsymbol{\beta}_k = (\beta_{k1}, \dots, \beta_{kp})^T \in \mathbb{R}^p.
\end{gather}
Here, $K$ is the number of neurons, $p$ is again the number of predictors, $j = 1, \dots, n$ with $n$ denoting the number of training observations, and $g(z) = \text{max(0, z)}$ is the ReLU function as usual. 

It has been mentioned that the posterior distributions of Bayesian neural networks are often multimodal as several equivalent combinations of parameters can result in the same prediction \citep{jospin2022hands}. However, for this illustration, this is not a concern as one is rather interested in the predictive performance. In that case, the most important parameter is $\sigma$ because whenever the posterior samples of $\sigma$ are close to the true value, the predictive error seems to be the smallest. To alleviate the issue of unidentifiability that could also impact the sampling of $\sigma$, it is assumed that $w_1 > 0$ and $w_2 > 0$ in the sampling procedure by rather defining $w_1^* = \log(w_1)$ and $w_2^* = \log(w_2)$. As before, $\gamma = \log(\sigma^2)$ is also introduced. This leads to $\mathbf{q} = (\gamma, \alpha, w_1^*, w_2^*, \delta_1, \delta_2, \boldsymbol{\beta}_1, \boldsymbol{\beta}_2)^T$ as the vector of interest. To simplify the setting, an exponential distribution with rate 1 is chosen as the prior for $\sigma$ while the remaining quantities have a standard normal distribution prior.  

\begin{figure} 
     \centering
     \includegraphics[width=0.9\textwidth]{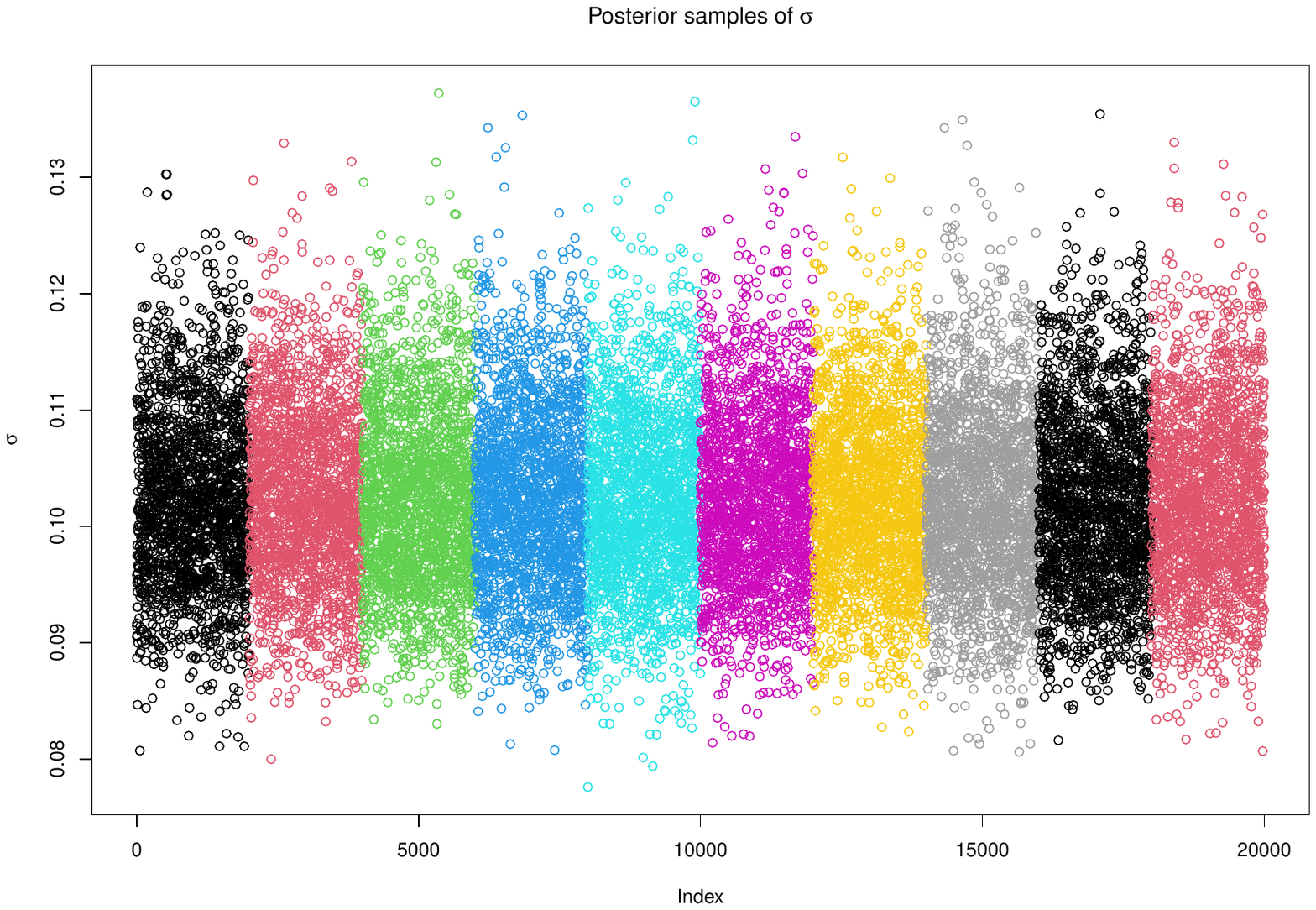}
     \caption{A plot of the obtained posterior samples from the 10 independent trajectories. Each color represents a distinct and independent trajectory targeting the presented Bayesian neural network model on the simulated data set. }
     \label{fig:bnn}
\end{figure}

For the purpose of illustration, both $K$ and $p$ are set to 2, i.e. a situation with two hidden neurons and two covariates is investigated. Also, let $\alpha = 0$, $\delta_1 = 0.5$, $\delta_2 = -0.5$, $\boldsymbol{\beta}_1 = (1, 0)^T$, $\boldsymbol{\beta}_1 = (-0.1, 1)^T$, $w_1 = w_2 = 1$, and $\sigma = 0.1$ be the true parameter values. A data set of $n = 100$ training observations is then simulated using the true parameter values. Table \ref{table:bnn} displays the result obtained by combining samples obtained from 10 independent trajectories. Here, the adaptive burn-in period lasts for 10000 time units for each trajectory before the sampling period with $T=100000$ and $N=100000$. One can observe that the posterior mean of $\sigma$ is close to the true value, indicating that the predictive performance of the model is adequate, and the methodology produces samples from the correct target. As a check, it is also seen that all trajectories produce comparable samples across the different trajectories from Figure \ref{fig:bnn}, where every 50th sample is plotted.

\begin{table}
    \centering
    \begin{tabular}{cccc}
         Parameter & Posterior Mean & Posterior SD  & 95\% Posterior Credible Interval \\
         \hline
         $\sigma$ & $0.102$ & $0.008$ & $(0.089, 0.119)$ \\
    \end{tabular}
    \caption{Results of the Bayesian neural network model based on 10 independent trajectories. }
    \label{table:bnn}
\end{table}

\begin{figure} 
     \centering
     \includegraphics[width=0.75\textwidth]{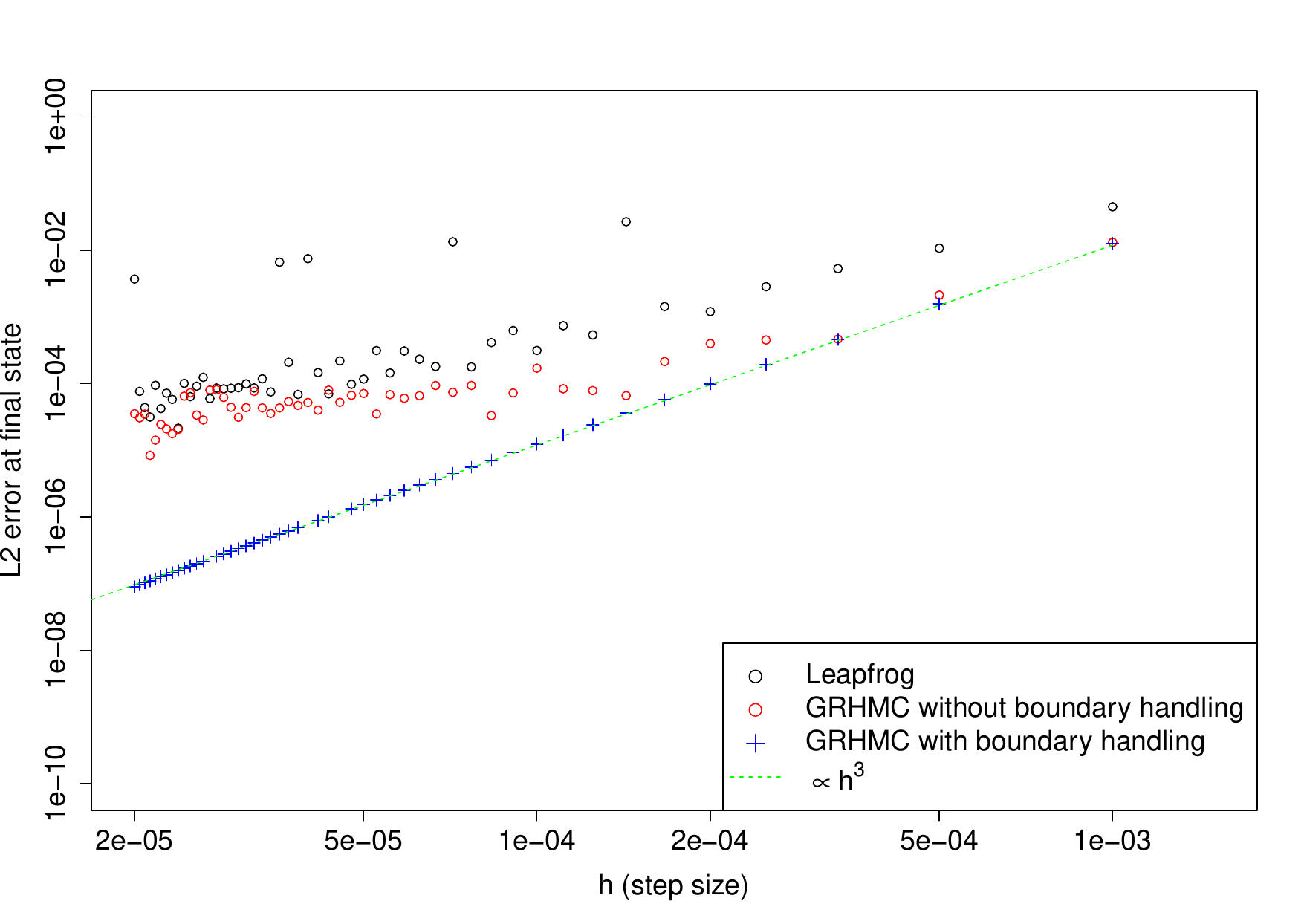}
     \caption{$L_2$-error of the final state $\mathbf{z}(T)=(\mathbf{q}(T), \mathbf{p}(T))$ for the Bayesian neural network setup against the step size $h$ in a log-log scale. The black dots are the errors obtained by the leapfrog method, the red points and blue crosses represent the errors from the GRHMC procedure without and with proper handling at the boundary of discontinuous gradient.}
     \label{fig:comparison_bnn_global_error}
\end{figure}

Finally, a similar procedure of checking the $L_2$-error as in Figure \ref{fig:comparison_global_error} is performed. $\mathbf{q}(0)$ is set to the true parameter values while $\mathbf{p}(0) = (0.1, -0.1, 0.1, -0.1, 0.1, -0.1, 0.1, -0.1, 0.1, -0.1)^T$. Here, the system is evolved for $T = 0.5$ time units. Again, one can observe from Figure \ref{fig:comparison_bnn_global_error} that the proposed method has the smallest error for a given value of $h$ when $h$ is small while preserving the correct integration order.  

\subsection{Regime switching volatility model} \label{sec:regime_switching}
In this final section, the proposed method from Section \ref{sec:piecewise_smooth} is applied to fit a regime switching volatility model on both a simulated and real data set. In econometrics, models involving state switching are common in order to capture the dynamics of price changes, where it is typical that the observed quantity cycles between periods with high volatility and time spans with less fluctuations \citep{chang2017new, chang2023oil}. Motivated by these applications, the following simplified model is examined in the current situation: 
\begin{equation} \label{eq:volatility_model_eq1}
    Y_t = \sigma_t \epsilon_t, 
    \, \sigma_t^2 = \begin{cases}
        \sigma_H^2, Z_t > 0, \\
        \sigma_L^2, \text{ otherwise},
    \end{cases} \;
    Z_t = Z_{t-1} + \eta_t, \, t = 1 \dots T 
\end{equation}
\begin{equation} \label{eq:volatility_model_eq2}
    \text{Corr}(\epsilon_t, \eta_t) = \rho, \;
    \eta_t \sim N(0, 1), \;
    \epsilon_t \sim N(0,1).
\end{equation}
In other words, the observation $Y_t$ is allowed to switch between two different states, a high volatility state denoted by $\sigma_h^2$ whenever the latent process $Z_t$ at the given time is above $0$ and a low volatility state represented by $\sigma_l^2$ if otherwise. Here, $Z_t$ follows a standard random walk. 
It is well known \citep[see e.g.][]{YU2005165} that certain financial time series are subject to the so-called leverage effects, where negative asset returns are associated with increases in volatility. In the present model given by \eqref{eq:volatility_model_eq1}-\eqref{eq:volatility_model_eq2}, such effects are captured by allowing $\eta_t$ and $\epsilon_t$ to be (generally negatively) correlated.

To perform a full Bayesian analysis under \eqref{eq:volatility_model_eq1}-\eqref{eq:volatility_model_eq2} requires sampling $\mathbf{q} = (Z_{1:T}, \rho, \sigma_L, \sigma_H)$ conditionally on the data after specifying priors for $\sigma_H$, $\sigma_L$, and $\rho$. Based on \eqref{eq:volatility_model_eq1}, it is clear that the target density is piecewise smooth as the density of $Y_t$ changes whenever $Z_t$ switches sign for a given $t$. Therefore, in order to sample from the model defined by \eqref{eq:volatility_model_eq1}-\eqref{eq:volatility_model_eq2} using Hamiltonian Monte Carlo methods, the proposed strategy from Section \ref{sec:piecewise_smooth} can be applied after similar transformations as before to map $\rho, \sigma_L$ and $\sigma_H$ to unconstrained domains. For $\sigma_L$ and $\sigma_H$, the transformation $\gamma = \log(\sigma^2)$ is therefore performed again. To map $\rho \in (-1, 1)$ to $\mathbb R$, the following transformation $\rho = \tanh(\rho^*), \, \rho^* \in \mathbb R$, is opted for here. Thus, the variables of interest for sampling purposes are $\mathbf{q} = (Z_{1:T}, \rho^*, \gamma_L, \gamma_H)$. Finally, to ensure that $\gamma_H > \gamma_L$, the methodology 
of GRHMC for restricted domain \citep{2311.14492} is also applied here. 

The presented model is fitted to a classical real data set consisting of Dollar/Pound exchange rates between October 1st 1981 and June 28th 1985 previously considered by e.g. \cite{harv:ruiz:shep:1994,shepard_pitt97,durbin2001tsa,liesenfeld_richard_06,KLEPPE20123105}. Here, the prior of $\rho$ is chosen to be a transformed beta distribution defined on the interval between -1 and 1 with both shape parameters equal to $2$. An exponential prior with rate parameter $1$ and $0.5$ is used for $\gamma_L$ and $\gamma_H$, respectively. 10 independent trajectories are simulated with $T = 100000$, $N = 100000$, $\lambda = 0.2$, $\mathbf m = \mathbf{0}_{T + 3}$, and $\mathbf S = \mathbf I_{T + 3}$, where the former half is discarded as burn-in samples. Note that the randomized sparse and deterministic reflection kernel will lead to the same momentum vector after reflection as all the normal vectors in this case only have one single non-zero component. Therefore, the former procedure is chosen in this case. 
\begin{table}
    \centering
    \begin{tabular}{ccccc}
         Parameter & Posterior Mean & Posterior Median & Posterior SD  & 95\% Posterior Credible Interval \\
         \hline
        $\rho$ & $-0.283$ & $-0.305$ & $0.212$ & $(-0.633, 0.179)$ \\
        $\sigma_L$ & $0.559$ & $0.560$ & $0.025$ & $(0.507, 0.605)$ \\
        $\sigma_H$ & $1.250$ & $1.241$ & $0.111$ & $(1.058, 1.490)$ \\
        \hline
    \end{tabular}
    \caption{Different results from the posterior samples of the volatility model for the real data based on 10 independent trajectories.}
    \label{table:volatility_real_data}
\end{table}
\begin{figure} 
     \centering
     \includegraphics[width=0.75\textwidth]{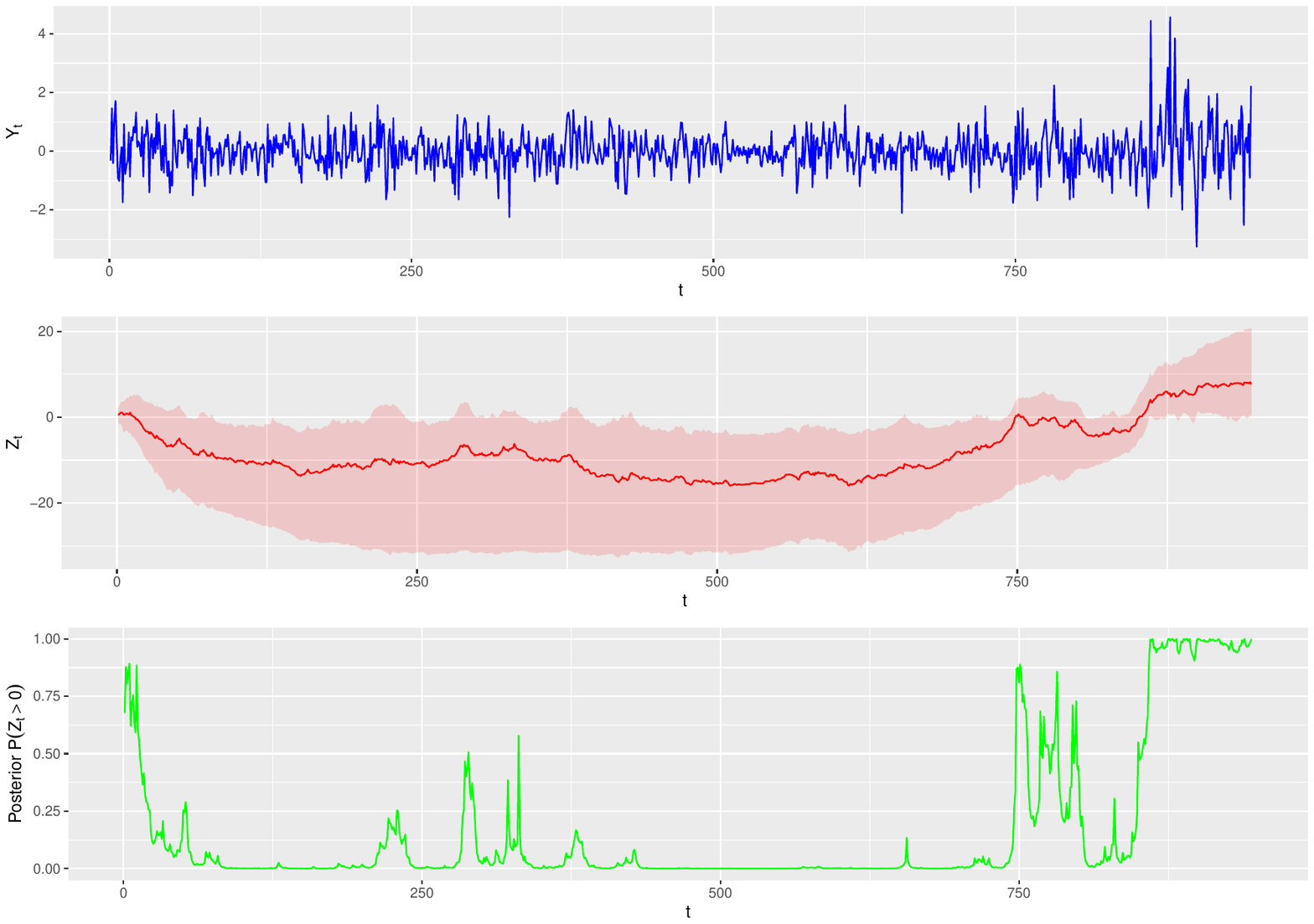}
     \caption{Results of the volatility model for the real data. Top plot: A plot of the observed quantity $Y_t$. Middle plot: A plot of the posterior mean of $Z_t$ with the 95\% credible interval. Bottom plot: A plot of the posterior probability of being in the high volatility state.}
     \label{fig:volatility_real_data}
\end{figure}
Interestingly, one can notice from the left plot of Figure \ref{fig:volatility_real_data} that the minor "spikes" at for instance the time periods around $t = 0$ and $t = 750$ are somewhat captured as high volatility situations based on the right plot. Similarly, during the time period close to $t = 500$ where the variances of the observations are smaller, the posterior probability of being in the high volatility state is essentially zero. Finally, at the end where the highest variation is observed, the same quantity tends towards one, almost guaranteeing that the observable variable is located in the high volatility state during this time period. Therefore, on the basis of Table \ref{table:volatility_real_data} and Figure \ref{fig:volatility_real_data}, it seems that the main features of the observations are reasonably captured by the given model. 

\section{Discussion} \label{sec:discussion}
In this project, Generalized Hamiltonian Monte Carlo processes intended for sampling of continuous target densities with discontinuous gradients and piecewise smooth densities have been presented. In both cases, the standard GRHMC is used to efficiently explore the interior of each region while specific updates are proceeded in order for the process to still admit the correct invariant distribution after boundary crossings. The proposed methods therefore combine both the strengths of Hamiltonian Monte Carlo methods with the possibilities of non-smooth targets. Thus, the contribution in this project can enhance further developments of realistic, but more complex models that in theory cannot be fitted using traditional gradient-based methods. Examples of challenging models that can be sampled with the introduced methodology have been demonstrated, with simulations verifying that the techniques work as intended. 

More work related to the proposed methodology is planned. In particular, automating the "integration beyond region" process, as explained in Section \ref{sec:numerical_implementation}, would require a specialized implementation to work using automatic differentiation software. This is currently being investigated within the context of the \texttt{pdmphmc} package (\url{https://torekleppe.github.io/pdmphmc.doc/}). In addition, the theoretical properties of the shrinkage prior in conjunction with the parameterization proposed in Section \ref{sec:reg_lin_reg} will be examined, and more specialized software based on the methodology currently proposed for this model will be examined.

\bibliographystyle{chicago}
\bibliography{main}
\newpage

\appendix
\begin{center}
\Large Supplementary material to "Numerical Generalized Randomized Hamiltonian Monte Carlo for piecewise smooth target densities" by Jimmy Huy Tran and Tore Selland Kleppe
\end{center}
\setcounter{page}{1}
\section{Time reversibility and volume preserving property of the overall Hamiltonian flow for continuous targets with discontinuous gradients} \label{app:time_reverse_overall_flow}
In this section, the arguments for why \eqref{eq:involution_overall_flow} and \eqref{eq:volume_preserving_overall_flow} are true will be presented. To show this, consider a general situation where evolving the state by $T$ times units from $\mathbf{\Bar{z}}(\tau)$ to $\varphi_{T \mid \mathbf{\Bar{z}}}(\mathbf{\Bar{z}} (\tau)) = \mathbf{\Bar{z}}(\tau + T)$ leads to $B$ boundary crossing events. Denote $\mathbf{\Bar{z}}_0 = \mathbf{\Bar{z}}(\tau)$, $\mathbf{\Bar{z}}_b, \, b = 1, \dots, B,$ as the states at these $B$ events and $T^*_b, \, b = 1, \dots, B,$ the time between such events (except for $b = 1$ where this corresponds to the time until the evolved system hits the boundary for the first time), $\mathbf{\Bar{z}}_{B + 1} = \mathbf{\Bar{z}}(\tau + T)$ and $T^*_{B + 1}$ the time from the state at the last boundary event until the total $T$ time units have been elapsed in total, i.e. $T^*_{B + 1} = T - \sum_{b=1}^B T_b^*$. Also, the following properties for the remaining parts are assumed, which can be intuitive when thinking of a path of a particle in a two-dimensional plane:
\begin{align}
    R(\mathbf{\Bar{z}}_{b}^-) &= R(\mathbf{\Bar{z}}_{b - 1}^+), \, b = 1, \dots, B + 1, \label{eq:time_reverse_overall_flow_nr1}  \\ 
    R(\mathbf{\Bar{z}}_{b}^-) &= R(\mathbf{\Bar{z}}_{b}^+), \, b = {0, \, B+1}, \label{eq:time_reverse_overall_flow_nr2} \\
    R((\mathbf R \circ \mathbf{\Bar{z}}_b)^+) &= R(\mathbf{\Bar{z}}_{b}^-), \, b = 1, \dots, B \label{eq:time_reverse_overall_flow_nr3}
\end{align}

Evolving the system from the initial state $\mathbf{\Bar{z}}_0$ to the state at the first boundary crossing event is essentially equivalent to $\mathbf{\Bar{z}}_1 = \varphi_{T^*_1, R(\mathbf{\Bar{z}}_0)}(\mathbf{\Bar{z}}_0) = \varphi_{T^*_1, R(\mathbf{\Bar{z}}_0^+)}(\mathbf{\Bar{z}}_0)$. Using \eqref{eq:flow_one_boundary_crossing_event} and \eqref{eq:flow_many_boundary_crossing_event}, one also observes that $\mathbf{\Bar{z}}_2 = \varphi_{T^*_2, R(\mathbf{\Bar{z}}_1^+)}(\mathbf{\Bar{z}}_1) = \varphi_{T^*_2, R(\mathbf{\Bar{z}}_1^+)} \circ \varphi_{T^*_1, R(\mathbf{\Bar{z}}_0^+)}(\mathbf{\Bar{z}}_0)$. Iteratively applying the same relations leads to
\begin{equation} \label{eq:time_reverse_overall_flow_nr4}
\begin{split}
    \mathbf{\Bar{z}}_{B + 1} = (\mathbf{\Bar{q}}_{B + 1}, \mathbf{\Bar{p}}_{B + 1}) &= \varphi_{T \mid \mathbf{\Bar{z}}}(\mathbf{\Bar{z}}_0) \\ 
    &= \varphi_{T^*_{B+1}, R(\mathbf{\Bar{z}}_{B}^+)} \circ \dots \circ \varphi_{T^*_1, R(\mathbf{\Bar{z}}_0^+)} (\mathbf{\Bar{z}}_0) \\
    &= \circ_{b = 1}^{B+1} \varphi_{T^*_{B+2-b}, R(\mathbf{\Bar{z}}_{B+1-b}^+)}(\mathbf{\Bar{z}}_0).
\end{split}
\end{equation} 
Now, set $\mathbf{\Tilde{\Bar{z}}}_0 = \mathbf R \circ_{b = 1}^{B+1} \varphi_{T^*_{B+2-b}, R(\mathbf{\Bar{z}}_{B+1-b}^+)}(\mathbf{\Bar{z}}_0) = \mathbf R \circ \mathbf{\Bar{z}}_{B + 1} = (\mathbf{\Bar{q}}_{B+1}, -\mathbf{\Bar{p}}_{B+1})$. Equation \eqref{eq:flow_many_boundary_crossing_event} implies
\begin{equation}\label{eq:time_reverse_overall_flow_nr5}
    \varphi_{T \mid \mathbf{\Bar{z}}}(\mathbf{\Tilde{\Bar{z}}}_0) = \varphi_{T - \Tilde{T}^*_1 \mid \mathbf{\Bar{z}}} \circ \varphi_{\Tilde{T}^*_1, R(\mathbf{\Tilde{\Bar{z}}}_0^+)} (\mathbf{\Tilde{\Bar{z}}}_0),
\end{equation}
where $\Tilde{T}^*_1$ is now the time from the new initial state $\mathbf{\Tilde{\Bar{z}}}_0$ until the evolved state arrives at a boundary point for the first time. However, note that 
\begin{equation} \label{eq:time_reverse_overall_flow_nr6}
\begin{split}
        \varphi_{\Tilde{T}^*_1, R(\mathbf{\Tilde{\Bar{z}}}_0^+)} (\mathbf{\Tilde{\Bar{z}}}_0)
        &= \varphi_{\Tilde{T}^*_1, R({\mathbf{\Bar{z}}}_{B+1}^-)} (\mathbf{\Tilde{\Bar{z}}}_0) \\
        &= \varphi_{{T}^*_{B+1}, R({\mathbf{\Bar{z}}}_{B}^+)} (\mathbf{\Tilde{\Bar{z}}}_0),
\end{split}
\end{equation}
where the first equality is a result of \eqref{eq:time_reverse_overall_flow_nr3} while \eqref{eq:time_reverse_overall_flow_nr1}, the fact mentioned regarding the time between boundary crossing events for \eqref{eq:flow_many_boundary_crossing_event} and reversibility of the region-specific flow lead to the last equality. Thus, 
again after using \eqref{eq:flow_one_boundary_crossing_event} and \eqref{eq:flow_many_boundary_crossing_event},  
\begin{equation} \label{eq:time_reverse_overall_flow_nr7}
    \begin{split}
        & \mathbf R \circ \varphi_{T - T^*_{B+1} \mid \mathbf{\Bar{z}}} \circ \varphi_{{T}^*_{B+1}, R({\mathbf{\Bar{z}}}_{B}^+)} \circ \mathbf R \circ \varphi_{T^*_{B+1}, R(\mathbf{\Bar{z}}_{B}^+)} \circ_{b = 2}^{B+1} \varphi_{T^*_{B+2-b}, R(\mathbf{\Bar{z}}_{B+1-b}^+)}(\mathbf{\Bar{z}}_0) \\
        &= \mathbf R \circ \varphi_{T - T^*_{B+1} \mid \mathbf{\Bar{z}}} \circ \mathbf R \circ_{b = 2}^{B+1} \varphi_{T^*_{B+2-b}, R(\mathbf{\Bar{z}}_{B+1-b}^+)}(\mathbf{\Bar{z}}_0)
    \end{split}
\end{equation}
due to \eqref{eq:involution_single_region_nr2} and noticing that $\mathbf R = \mathbf R^{-1}$.
\newline

In a similar manner, define $\mathbf{\Tilde{\Bar{z}}}_1 = \mathbf R \circ_{b = 2}^{B+1} \varphi_{T^*_{B+2-B}, R(\mathbf{\Bar{z}}_{B+1-b}^+)}(\mathbf{\Bar{z}}_0) = \mathbf R \circ \mathbf{\Bar{z}}_B$. Equation \eqref{eq:flow_many_boundary_crossing_event} gives
\begin{equation} \label{eq:time_reverse_overall_flow_nr8}
    \varphi_{T - T^*_{B+1} \mid \mathbf{\Bar{z}}} (\mathbf{\Tilde{\Bar{z}}}_1) = \varphi_{T - T^*_{B+1} - \Tilde{T}^*_2 \mid \mathbf{\Bar{z}}} \circ \varphi_{\Tilde{T}^*_2, R(\mathbf{\Tilde{\Bar{z}}}_1^+)} (\mathbf{\Tilde{\Bar{z}}}_1).
\end{equation}
Again, the conditions from \eqref{eq:time_reverse_overall_flow_nr1} and \eqref{eq:time_reverse_overall_flow_nr3} yield 
\begin{equation} \label{eq:time_reverse_overall_flow_nr9}
    \begin{split}
        \varphi_{\Tilde{T}^*_2, R(\mathbf{\Tilde{\Bar{z}}}_1^+)} (\mathbf{\Tilde{\Bar{z}}}_1) &= \varphi_{\Tilde{T}^*_2, R({\mathbf{\Bar{z}}}_B ^-)} (\mathbf{\Tilde{\Bar{z}}}_1) \\
        &= \varphi_{T^*_B, R({\mathbf{\Bar{z}}}_{B-1}^+)} (\mathbf{\Tilde{\Bar{z}}}_1).
    \end{split}
\end{equation}
Equation \eqref{eq:time_reverse_overall_flow_nr7} can therefore be rewritten as 
\begin{equation} \label{eq:time_reverse_overall_flow_nr10}
\begin{split}
    &\mathbf R \circ \varphi_{T - \sum_{b' = 1}^2 T^*_{B+2-b'} \mid \mathbf{\Bar{z}}} \circ \varphi_{T^*_B, R({\mathbf{\Bar{z}}}_{B-1}^+)} \circ \mathbf R \circ \varphi_{T^*_{B}, R(\mathbf{\Bar{z}}_{B-1}^+)} \circ_{b = 3}^{B+1} \varphi_{T^*_{B+2-b}, R(\mathbf{\Bar{z}}_{B+1-b}^+)}(\mathbf{\Bar{z}}_0) \\
    &= \mathbf R \circ \varphi_{T - \sum_{b' = 1}^2 T^*_{B+2-b'} \mid \mathbf{\Bar{z}}} \circ \mathbf R \circ_{b = 3}^{B+1} \varphi_{T^*_{B+2-b}, R(\mathbf{\Bar{z}}_{B+1-b}^+)}(\mathbf{\Bar{z}}_0).
\end{split}
\end{equation}
Running through the same arguments $B$ times in total, one arrives at
\begin{equation}
    \mathbf R \circ \varphi_{T^*_1, R(\mathbf{\Bar{z}}_0 ^+)}  \circ \mathbf R \circ \varphi_{T^*_1, R(\mathbf{\Bar{z}}_0^+)}.
\end{equation}
But since this corresponds to a specific region-flow, the result is simply an identity operator due to \eqref{eq:flow_many_boundary_crossing_event}. Hence, equation \eqref{eq:involution_overall_flow} also holds. 

For the volume preservation property, one can start from \eqref{eq:time_reverse_overall_flow_nr4}. Since the overall flow is a composition of several flows, the multivariate chain rule states that the Jacobian matrix of the overall flow is simply the product of the Jacobian matrices of all composed flows: 
\begin{equation}
    \nabla_{\mathbf{\Bar{z}}_0}\varphi_{T \mid \mathbf{\Bar{z}}}(\mathbf{\Bar{z}}_0) = \squarebrackets{\nabla_{\mathbf{\Bar{z}}_B} \varphi_{T^*_{B+1}, R(\mathbf{\Bar{z}}_{B}^+)}(\mathbf{\Bar{z}}_B)} \cdots  \squarebrackets{\nabla_{\mathbf{\Bar{z}}_1} \varphi_{T^*_2, R(\mathbf{\Bar{z}}_1^+)}(\mathbf{\Bar{z}}_1)} \squarebrackets{\nabla_{\mathbf{\Bar{z}}_0} \varphi_{T^*_1, R(\mathbf{\Bar{z}}_0^+)} (\mathbf{\Bar{z}}_0)}
\end{equation}
The determinant of the desired Jacobian on the left-hand side is therefore also the product of the determinant of each of the specific region flow:
\begin{equation}
    \text{det}\parentheses{\nabla_{\mathbf{\Bar{z}}_0}\varphi_{T \mid \mathbf{\Bar{z}}}(\mathbf{\Bar{z}}_0)} = \text{det}\parentheses{\nabla_{\mathbf{\Bar{z}}_B} \varphi_{T^*_{B+1}, R(\mathbf{\Bar{z}}_{B}^+)}(\mathbf{\Bar{z}}_B)} \cdots  \text{det}\parentheses{\nabla_{\mathbf{\Bar{z}}_1} \varphi_{T^*_2, R(\mathbf{\Bar{z}}_1^+)}(\mathbf{\Bar{z}}_1)} \text{det}\parentheses{\nabla_{\mathbf{\Bar{z}}_0} \varphi_{T^*_1, R(\mathbf{\Bar{z}}_0^+)} (\mathbf{\Bar{z}}_0)}
\end{equation}
But again, since each of the specific region flow follows the standard Hamiltonian dynamics, the determinants on the right-hand side are either $1$ or $-1$. Thus, the determinant of the Jacobian of the overall flow must be either $1$ or $-1$, which implies that \eqref{eq:flow_many_boundary_crossing_event} is also volume preserving. 

\section{Randomized reflection kernel leads to correct invariant distribution for piecewise smooth target densities} \label{app:show_randomized_reflection_piecewise}
It is mentioned that the boundary transition kernel deduced by the modified reflection and refraction process satisfies \eqref{eq:invariant_boundary_kernel} for situations with target densities following \eqref{eq:piecewise_smooth_density}. 

First, notice if the reflection and refraction transition kernel $K$ admits \eqref{eq:randomized_reflection} for the reflection part, then $K'$ becomes 
\begin{equation} \label{eq:show_randomized_reflection_piecewise_nr1}
    \mathbf{\Bar{p}}' = \begin{cases}
        -\mathbf{\Bar{p}} + \parentheses{ \sqrt{(\mathbf{\Bar{p}}^T \mathbf{\hat{n}}(\mathbf{\Bar{q}}))^2 + 2 \Delta U} + \mathbf{\Bar{p}}^T \mathbf{\hat{n}}(\mathbf{\Bar{q}})} \mathbf{\hat{n}}(\mathbf{\Bar{q}}), \text{ if } \mathbf{\Bar{p}} \in  \mathcal{P}_\mathbf{\Bar{q}}^-, \\
        -\mathbf{\Bar{p}} + \parentheses{- \sqrt{(\mathbf{\Bar{p}}^T \mathbf{\hat{n}}(\mathbf{\Bar{q}}))^2 - 2 \Delta U} + \mathbf{\Bar{p}}^T \mathbf{\hat{n}}(\mathbf{\Bar{q}})} \mathbf{\hat{n}}(\mathbf{\Bar{q}}), \text{ if } \mathbf{\Bar{p}} \in  \mathcal{P}_\mathbf{\Bar{q}}^+ \text{ and } (\mathbf{\Bar{p}}^T \mathbf{\hat{n}}(\mathbf{\Bar{q}}))^2 > 2 \Delta U,\\
        \mathbf{x} - \parentheses{(-\mathbf{\Bar{p}} + \mathbf{x})^T \mathbf{\hat{n}}(\mathbf{\Bar{q}})} \mathbf{\hat{n}}(\mathbf{\Bar{q}}), \, \mathbf{x} \sim N(\mathbf{0}_d, \mathbf{I}_d) , \text{ if } \mathbf{\Bar{p}} \in  \mathcal{P}_\mathbf{\Bar{q}}^+ \text{ and } (\mathbf{\Bar{p}}^T \mathbf{\hat{n}}(\mathbf{\Bar{q}}))^2 < 2 \Delta U.\\
    \end{cases}
\end{equation}
Here, the fact that $\mathbf{\Bar{p}} \in  \mathcal{P}_\mathbf{\Bar{q}}^-$ implies $-\mathbf{\Bar{p}} \in  \mathcal{P}_\mathbf{\Bar{q}}^+$, and vice versa, is applied. The first two situations are exactly the same as in \citet{chevallier2021pdmp}, hence the proof for the former two is identical as in the referred article. Therefore, it remains to check the reflection part. Also, observe that $\mathbf{\Bar{p}}^T \mathbf{\hat{n}}(\mathbf{\Bar{q}}) = \mathbf{\Bar{p}}'^T \mathbf{\hat{n}}(\mathbf{\Bar{q}})$, thus $\mathbf{\Bar{p}}' \in \mathcal{P}_\mathbf{\Bar{q}}^+ $. Accordingly, condition \eqref{eq:invariant_boundary_kernel} becomes 
\begin{gather}
    m_\mathbf{\Bar{q}}(\mathbf{\Bar{p}}') = \int_{\mathcal{P}_\mathbf{\Bar{q}}^+} K'_\mathbf{\Bar{q}}(\mathbf{\Bar{p}}' \mid \mathbf{\Bar{p}}) m_\mathbf{\Bar{q}}(\mathbf{\Bar{p}}) d\mathbf{\Bar{p}} \label{eq:show_randomized_reflection_piecewise_nr2}, \\ 
    |\mathbf{\Bar{p}}'^T \mathbf{\hat{n}}(\mathbf{\Bar{q}})| p(\mathbf{\Bar{p}}') \pi_{k_2(\mathbf{\Bar{q}})}(\mathbf{\Bar{q}}) = \int_{\mathcal{P}_\mathbf{\Bar{q}}^+} K'_\mathbf{\Bar{q}}(\mathbf{\Bar{p}}' \mid \mathbf{\Bar{p}}) |\mathbf{\Bar{p}}^T \mathbf{\hat{n}}(\mathbf{\Bar{q}})| p(\mathbf{\Bar{p}}) \pi_{k_2(\mathbf{\Bar{q}})}(\mathbf{\Bar{q}}) d\mathbf{\Bar{p}} \label{eq:show_randomized_reflection_piecewise_nr3}, \\ 
    p(\mathbf{\Bar{p}}') = \int_{\mathcal{P}_\mathbf{\Bar{q}}^+} K'_\mathbf{\Bar{q}}(\mathbf{\Bar{p}}' \mid \mathbf{\Bar{p}}) p(\mathbf{\Bar{p}}) d\mathbf{\Bar{p}} \label{eq:show_randomized_reflection_piecewise_nr4}.
\end{gather}

To show that \eqref{eq:invariant_boundary_kernel} is true for the randomized reflection kernel is therefore equivalent to show that \eqref{eq:show_randomized_reflection_piecewise_nr4} holds. This can be done either by observing that the momentum update given in the latter case of \eqref{eq:show_randomized_reflection_piecewise_nr1} is an affine transformation of a standard normal distribution similar to \citet{kleppe2023} or explicitly computing the integral on the right-hand side of \eqref{eq:show_randomized_reflection_piecewise_nr4} (see e.g. \citet{wu2017generalized} for a similar approach). Either way, notice that 
\begin{align}
     \mathbf{\Bar{p}}' &= \mathbf{x} - \parentheses{(-\mathbf{\Bar{p}} + \mathbf{x})^T \mathbf{\hat{n}}(\mathbf{\Bar{q}})} \mathbf{\hat{n}}(\mathbf{\Bar{q}}) \\
     &= \mathbf{x} - (\mathbf{x}^T \mathbf{\hat{n}}(\mathbf{\Bar{q}})) \mathbf{\hat{n}}(\mathbf{\Bar{q}}) + (\mathbf{\Bar{p}}^T \mathbf{\hat{n}}(\mathbf{\Bar{q}})) \mathbf{\hat{n}}(\mathbf{\Bar{q}}).
\end{align}
The first two terms together represent the projection of $\mathbf{x}$ onto the $(d-1)$-dimensional space orthogonal to the normal vector. The last term corresponds to the component parallel to the normal vector. For simplicity, the former is considered first here. 

Equation \eqref{eq:show_randomized_reflection_piecewise_nr4} states that if the momentum vector just before reflection is distributed according to $p$, the corresponding reflected momentum vector should follow the same distribution. Note that the above expression of the randomized reflected momentum can also be written as
\begin{equation}
    \mathbf{\Bar{p}}' = \begin{bmatrix}
        \mathbf{\hat{n}}(\mathbf{\Bar{q}})\mathbf{\hat{n}}(\mathbf{\Bar{q}})^T & \mathbf{I}_d - \mathbf{\hat{n}}(\mathbf{\Bar{q}})\mathbf{\hat{n}}(\mathbf{\Bar{q}})^T
    \end{bmatrix} \begin{bmatrix}
        \mathbf{\Bar{p}} \\
        \mathbf{{x}}
    \end{bmatrix}.
\end{equation}
Under the GRHMC framework, $\mathbf{\Bar{p}}$ is simply a $d$-dimensional standard normal distributed variable such that $ \squarebrackets{\mathbf{\Bar{p}}^T, \mathbf{{x}}^T} \sim N(\mathbf{0}
_{2d}, \mathbf{I}_{2d})$. Consequently, the resulting vector $\mathbf{\Bar{p}}'$ is an affine transformation of a normally distributed variable, which implies that it is also normal. It therefore only remains to check that the covariance matrix is the identity matrix. Trivial computations yield
\begin{equation}
    \begin{split}
        \text{Var}(\mathbf{\Bar{p}}') &= \begin{bmatrix}
        \mathbf{\hat{n}}(\mathbf{\Bar{q}})\mathbf{\hat{n}}(\mathbf{\Bar{q}})^T & \mathbf{I}_d - \mathbf{\hat{n}}(\mathbf{\Bar{q}})\mathbf{\hat{n}}(\mathbf{\Bar{q}})^T
    \end{bmatrix} 
    \text{Var}\parentheses{\begin{bmatrix}
        \mathbf{\Bar{p}} \\
        \mathbf{{z}}
    \end{bmatrix}}
    \begin{bmatrix}
        \mathbf{\hat{n}}(\mathbf{\Bar{q}})\mathbf{\hat{n}}(\mathbf{\Bar{q}})^T & \mathbf{I}_d - \mathbf{\hat{n}}(\mathbf{\Bar{q}})\mathbf{\hat{n}}(\mathbf{\Bar{q}})^T
    \end{bmatrix} ^T \\
    &= \begin{bmatrix}
        \mathbf{\hat{n}}(\mathbf{\Bar{q}})\mathbf{\hat{n}}(\mathbf{\Bar{q}})^T & \mathbf{I}_d - \mathbf{\hat{n}}(\mathbf{\Bar{q}})\mathbf{\hat{n}}(\mathbf{\Bar{q}})^T
    \end{bmatrix} \begin{bmatrix}
        \mathbf{\hat{n}}(\mathbf{\Bar{q}})\mathbf{\hat{n}}(\mathbf{\Bar{q}})^T & \mathbf{I}_d - \mathbf{\hat{n}}(\mathbf{\Bar{q}})\mathbf{\hat{n}}(\mathbf{\Bar{q}})^T
    \end{bmatrix} ^T \\
    &= \mathbf{\hat{n}}(\mathbf{\Bar{q}})\mathbf{\hat{n}}(\mathbf{\Bar{q}})^T \mathbf{\hat{n}}(\mathbf{\Bar{q}})\mathbf{\hat{n}}(\mathbf{\Bar{q}})^T + \parentheses{\mathbf{I}_d - \mathbf{\hat{n}}(\mathbf{\Bar{q}})\mathbf{\hat{n}}(\mathbf{\Bar{q}})^T} \parentheses{\mathbf{I}_d - \mathbf{\hat{n}}(\mathbf{\Bar{q}})\mathbf{\hat{n}}(\mathbf{\Bar{q}})^T}^T \\
    &= \mathbf{\hat{n}}(\mathbf{\Bar{q}})\mathbf{\hat{n}}(\mathbf{\Bar{q}})^T + \mathbf{I}_d  - \mathbf{\hat{n}}(\mathbf{\Bar{q}})\mathbf{\hat{n}}(\mathbf{\Bar{q}})^T - \mathbf{\hat{n}}(\mathbf{\Bar{q}})\mathbf{\hat{n}}(\mathbf{\Bar{q}})^T + \mathbf{\hat{n}}(\mathbf{\Bar{q}})\mathbf{\hat{n}}(\mathbf{\Bar{q}})^T \mathbf{\hat{n}}(\mathbf{\Bar{q}})\mathbf{\hat{n}}(\mathbf{\Bar{q}})^T \\
    &= \mathbf{I}_d,
    \end{split}
\end{equation}
where the fact that $\mathbf{\hat{n}}(\mathbf{\Bar{q}})^T\mathbf{\hat{n}}(\mathbf{\Bar{q}}) = 1$ and $\parentheses{\mathbf{\hat{n}}(\mathbf{\Bar{q}})\mathbf{\hat{n}}(\mathbf{\Bar{q}})^T}^T = \mathbf{\hat{n}}(\mathbf{\Bar{q}})\mathbf{\hat{n}}(\mathbf{\Bar{q}})^T$ have been used in the last three lines. 

For the second approach, due to the randomness added by $\mathbf{x}$, one can safely assume that $\mathcal{P}_\mathbf{\Bar{q}}^+ = \mathbb{R}^d$. Based on this observation, it is convenient to perform a change of basis so that e.g. the first component of the transformed $\mathbf{\Bar{p}}$, denoted as $\mathbf{\Bar{p}}^{*} = (\Bar{p}_1^*, \cdots, \Bar{p}^*_d)^T$, points along the normal vector such that the remaining components are pointing along the space defined by the $(d-1)$ basis vectors that are orthogonal to the normal vector. In that case, note that one can rewrite the randomized part as
\begin{equation}
    \mathbf{x} - (\mathbf{x}^T \mathbf{\hat{n}}(\mathbf{\Bar{q}})) \mathbf{\hat{n}}(\mathbf{\Bar{q}}) = \parentheses{\mathbf{I}_d - \mathbf{\hat{n}}(\mathbf{\Bar{q}}) \mathbf{\hat{n}}(\mathbf{\Bar{q}})^T} \mathbf{x}. 
\end{equation}
Let $\boldsymbol{\Sigma}^* = \parentheses{\mathbf{I}_d - \mathbf{\hat{n}}(\mathbf{\Bar{q}}) \mathbf{\hat{n}}(\mathbf{\Bar{q}})^T}$. It can be shown that this matrix has a single eigenvalue equal to $0$ with the corresponding normalized eigenvector $\mathbf{Q}_1$ being the unit normal vector $\mathbf{\hat{n}}(\mathbf{\Bar{q}})$. This direction is represented in the new basis by the component $\Bar{p}_1^*$. The remaining normalized eigenvectors will then span the orthogonal projection space of the normal vector. Next, the eigendecomposition can be applied for $\boldsymbol{\Sigma}^*$. This yields
\begin{equation}
    \boldsymbol{\Sigma}^* = \mathbf{Q} \mathbf{D} \mathbf{Q}^T,
\end{equation}
where 
$\mathbf{D} = \text{diag}(0, \mathbf{1}_{d-1})$ so that 
\begin{equation}
    \mathbf{Q} = \begin{bmatrix}
\mathbf{Q}_1 & \mathbf{Q}_2 & ... & \mathbf{Q}_d
\end{bmatrix}
\end{equation}
is the matrix with the corresponding normalized eigenvectors.
To obtain $\mathbf{Q}_j, \, j = 2 \dots d$, it can be shown that $\Tilde{\mathbf{Q}}_j = (-n_j/n_1, \mathbf{0}_{d-1})^T + \mathbf{e}_j^T$ is an eigenvector of the eigenvalue equal to 1. Here, $\mathbf{e}_j$ denotes the vector with the value 1 in the $j$-th coordinate and 0 otherwise. Performing the Gram-Schmidt process on the following set of eigenvectors of the eigenvalue 1 $\curlybrackets{\Tilde{\mathbf{Q}}_2, \dots, \Tilde{\mathbf{Q}}_d}$ leads to  the orthonormal set of eigenvectors $\curlybrackets{{\mathbf{Q}}_2, \dots, {\mathbf{Q}}_d}$ for the corresponding eigenvalue with 
\begin{align}
    {\mathbf{Q}}_{jl} &= \begin{cases}
         -\frac{n_l n_j}{\sqrt{\sum_{j'=1}^{j-1} n_{j'}^2}\sqrt{\sum_{j'=1}^{j} n_{j'}^2}},  &l < j, \\
        \frac{\sqrt{\sum_{j'=1}^{j-1} n_{j'}}}{\sqrt{\sum_{j'=1}^{j} n_{j'}}}, &l = j, \\
        0, &l > j.
    \end{cases}
\end{align}
Since $\curlybrackets{{\mathbf{Q}}_1, \dots, {\mathbf{Q}}_d}$ is now a set of orthonormal vectors, $\mathbf{Q} \mathbf{Q}^T = \mathbf{I}_d$ so that transforming from $\mathbf{\Bar{p}}$ to $\mathbf{\Bar{p}}^* = \mathbf{Q}\mathbf{\Bar{p}}$ leads to $p(\mathbf{\Bar{p}}^*)=\mathcal{N}(\mathbf{\Bar{p}}^* \mid \mathbf{0}_d, \mathbf{I}_d)$. In addition, the determinant of the Jacobian of the transformation is simply 1. Therefore, condition \eqref{eq:show_randomized_reflection_piecewise_nr4} becomes 
\begin{equation} \label{eq:show_randomized_reflection_piecewise_nr5}
    p(\mathbf{\Bar{p}}^{*'}) = \int_{\mathcal{P}_\mathbf{\Bar{q}}^+} K'_\mathbf{\Bar{q}}(\mathbf{\Bar{p}}^{*} \mid \mathbf{\Bar{p}^*}) p(\mathbf{\Bar{p}}^*) d\mathbf{\Bar{p}}^*
\end{equation}
in the new basis defined by the normalized eigenvectors of $\boldsymbol{\Sigma}^*$.  

Using the results from above, one can now also argue why $\parentheses{\mathbf{I}_d - \mathbf{\hat{n}}(\mathbf{\Bar{q}}) \mathbf{\hat{n}}(\mathbf{\Bar{q}})^T} \mathbf{x}$ can be represented by a $(d-1)$-dimensional standard normal distribution. Since the space orthogonal to the normal vector is spanned by $\mathbf{Q}_2, \cdots, \mathbf{Q}_d$, consider ${W}_j = \mathbf{Q}_j^T \mathbf{x}, \,  j = 2, \cdots, d$, i.e. $\mathbf{x}$ is projected onto a given basis vector that spans the orthogonal projection space. Then, $W_j \sim N(0, 1)$ and $\text{Cov}(W_j, W_j')=0$ if $j \neq j'$.  Hence, in this new coordinate system, the randomized reflection kernel from \eqref{eq:show_randomized_reflection_piecewise_nr1} can now be expressed as 
\begin{equation}
    K'_\mathbf{\Bar{q}}(\mathbf{\Bar{p}}^{*'} \mid \mathbf{\Bar{p}^*}) = \delta_{\Bar{p}_1^*} (\Bar{p}_1^{*'}) \mathcal{N}(\Bar{p}_2^{*'} \cdots \Bar{p}_d^{*'} \mid \mathbf{0}_{d - 1}, \mathbf{I}_{d - 1}). 
\end{equation}
Inserting back to the right-hand side of \eqref{eq:show_randomized_reflection_piecewise_nr5} above yields
\begin{equation}
    \begin{split}
        \int_{\mathcal{P}_\mathbf{\Bar{q}}^+} K'_\mathbf{\Bar{q}}(\mathbf{\Bar{p}}^{*'} \mid \mathbf{\Bar{p}^*}) p(\mathbf{\Bar{p}}^*) d\mathbf{\Bar{p}}^* &= \int_{(\Bar{p}_1^*, \cdots, \Bar{p}_d^*)^T \in \mathbb{R}^d} 
        \delta_{\Bar{p}_1^*} (\Bar{p}_1^{*'}) \mathcal{N}(\Bar{p}_2^{*'}, \cdots, \Bar{p}_d^{*'} \mid \mathbf{0}_{d - 1}, \mathbf{I}_{d - 1})
        \mathcal{N}(\Bar{p}_1^{*}, \cdots, \Bar{p}_d^{*} \mid \mathbf{0}_{d}, \mathbf{I}_{d}) \, d\Bar{p}_1^{*} \cdots \Bar{p}_d^{*} \\
        &= \int_{(\Bar{p}_2^*, \cdots, \Bar{p}_d^*)^T \in \mathbb{R}^{d - 1}}
        \mathcal{N}(\Bar{p}_2^{*'}, \cdots, \Bar{p}_d^{*'} \mid \mathbf{0}_{d - 1}, \mathbf{I}_{d - 1})
        \mathcal{N}(\Bar{p}_1^{*'}, \cdots, \Bar{p}_d^{*} \mid \mathbf{0}_{d}, \mathbf{I}_{d}) \, d\Bar{p}_2^{*} \cdots \Bar{p}_d^{*} \\
        &= \mathcal{N}(\Bar{p}_1^{*'}, \cdots, \Bar{p}_d^{*'} \mid \mathbf{0}_{d}, \mathbf{I}_{d})
        \int_{(\Bar{p}_2^*, \cdots, \Bar{p}_d^*)^T \in \mathbb{R}^{d - 1}} 
        \mathcal{N}(\Bar{p}_2^{*}, \cdots, \Bar{p}_d^{*} \mid \mathbf{0}_{d-1}, \mathbf{I}_{d-1}) \, d\Bar{p}_2^{*} \cdots \Bar{p}_d^{*} \\
        &= \mathcal{N}(\Bar{p}_1^{*'}, \cdots, \Bar{p}_d^{*'} \mid \mathbf{0}_{d}, \mathbf{I}_{d}) = p(\mathbf{\Bar{p}}^{*'}).
    \end{split}
\end{equation}
This concludes the proof as the randomized reflection kernel does indeed satisfy \eqref{eq:show_randomized_reflection_piecewise_nr5} in the new basis, which in turn means that \eqref{eq:show_randomized_reflection_piecewise_nr4} is true in the original standard basis.  

\section{Tuning momentum refresh event rate} \label{app:tuning_lambda}
In many situations, it is customary to adapt the momentum refresh event rate $\lambda$ based on the dynamics of the problem at hand. A common quantity used for such purposes is the so-called "U-turn" time \citep{hoffman2014no}. Without loss of generality, assume that time is reset to zero whenever an event occurs, i.e. $\mathbf{\Bar{q}}(0)$ represents the position right after a momentum refresh. The "U-turn time" $\omega$ from the given position is given as 
\begin{equation} \label{eq:u_turn_condition}
    \omega(\mathbf{\Bar{z}}(0)) = \text{inf}\curlybrackets{t > 0: (\mathbf{\Bar{q}}(t) - \mathbf{\Bar{q}}(0))^T \mathbf{\Bar{p}}(t) < 0}. 
\end{equation}
\citet{kleppe2022connecting} proposed a tuning approach of $\lambda$ based on the quantity above by applying exponential moving average on the computed values of $\omega$. In case $\omega$ is larger than the next momentum refresh event time, one needs to perform the numerical integration further to locate $\omega$ before discarding this part of the dynamics. Therefore, an alternative approach is opted for in this project by assuming that $\omega$ follows an exponential distribution with rate parameter $\lambda$. If $\tau_i$ denotes the time between the $(i-1)$-th and $i$-th refresh event, let $\omega^*_i = \text{min}(\omega_i, \tau_i)$ be the observed U-turn time, and $U_i = 1$ if $\omega^*_i = \omega_i$. This is the same as assuming that the U-turn time in this part of the trajectory is censored if it is not observed before the proceeding momentum refresh event time. Therefore, at the momentum refresh event number $u^*$, $\lambda$ is updated according to the maximum likelihood estimate with censoring:
\begin{equation}
    \lambda = \frac {\sum_{i=1}^{u^*} U_i}  {\sum_{i=1}^{u^*} \omega^*_i}
\end{equation}

The strategy above works numerically fine in the case with no reflection. On the other hand, if reflections appear like in the GRHMC processes for piecewise smooth targets, the numerical root finder might indicate that another root has been found immediately due to the U-turn condition. However, when the procedure evaluates all the root functions to check which one is closest to zero in absolute value, the root function regarding the boundary collision could still be returned due to numerical precision. This leads to a double reflection, and the component of the momentum vector parallel to the normal vector reverses back. To alleviate this problem, one could either assume that the U-turn time is censored so that the U-turn condition is reset with a new initial $\Bar{\mathbf{q}}$ or that the time between two consecutive events of either momentum refresh or reflection is also included as a part of the estimation of $\lambda$. 
For the current purpose, the former approach is resorted to in order to prevent the trajectories inheriting a random walk behaviour with the much larger estimate of momentum refresh event rate. Similar to standard momentum refresh, $\lambda$ will also be updated after each reflection during the adaptive burn-in period. Finally, if the situation is such that no U-turn events are detected, the momentum refresh event rate is set to a prespecified minimum value $\lambda_\text{min}$. 

\section{Additional details regarding the regularized linear regression model} \label{app:details_reg_lin_reg}

In this section of the Appendix, additional details related to the model presented in Section \ref{sec:reg_lin_reg} are presented. It has been mentioned that the shrinkage effect is determined by the specified combination of prior $P(\beta_i = 0$) and $\text{Var}(\beta_i \mid \beta_i \neq 0)$, which are again determined by the prior hyperparameters $\mu$ and $\rho$. The goal here is therefore to establish the equations needed to calculate $\mu$ and $\rho$ based on the desired prior $P(\beta_i = 0$) and $\text{Var}(\beta_i \mid \beta_i \neq 0)$. 

\subsection{Relationship between the prior parameters, inclusion probability and conditional variance}

First, by \eqref{eq:beta_wrt_beta_plus_and_minus}, notice that the prior unconditional expectation of $\beta_i$ is 
\begin{equation}
\begin{split}
    \mathbb{E}(\beta_i)&=\mathbb{E}(\text{max}(0, \beta_i^+)) - \mathbb{E}(\text{max}(0, \beta_i^-)) \\
    &= \integral{-\infty}{\infty} \text{max}(0, \beta_i^+) \mathcal{N}(\beta_i^+ \mid \mu, \rho^2) d\beta_i^+ - \integral{-\infty}{\infty} \text{max}(0, \beta_i^-) \mathcal{N}(\beta_i^- \mid \mu, \rho^2) d\beta_i^- \\
    &= \integral{0}{\infty} \beta_i^+ \mathcal{N}(\beta_i^+ \mid \mu, \rho^2) d\beta_i^+ - \integral{0}{\infty} \beta_i^- \mathcal{N}(\beta_i^- \mid \mu, \rho^2) d\beta_i^- \\
    &= \frac{\rho \exp(-\frac{\mu^2}{2\rho^2})}{\sqrt{2 \pi}} + \mu \frac{\text{erf}(\frac{\mu}{\sqrt{2\pi}\rho}) + 1}{2} - \parentheses{\frac{\rho \exp(-\frac{\mu^2}{2\rho^2})}{\sqrt{2 \pi}} + \mu \frac{\text{erf}(\frac{\mu}{\sqrt{2\pi}\rho}) + 1}{2}} \\
    &= \rho \phi(\mu / \rho) + \mu \Phi(\mu / \rho) - (\rho \phi(\mu / \rho) + \mu \Phi(\mu / \rho)) \\ 
    &= 0. 
\end{split}
\end{equation}
Here, $\parentheses{\text{erf}(\frac{\mu}{\sqrt{2\pi}\rho}) + 1} / 2 = \Phi(\mu / \rho)$ and $\exp(-\mu^2/(2\rho^2)) / \sqrt{2 \pi} = \phi(\mu / \rho)$, where $\Phi$ and $\phi$ represent the CDF and PDF of $N(0, 1)$, respectively. For later calculations, it is convenient to define $\mu_1^* := \mathbb{E}(\text{max}(0, \beta_i^+)) = \mathbb{E}(\text{max}(0, \beta_i^-)) = \rho \phi(\mu / \rho) + \mu \Phi(\mu / \rho)$. 

Next, due to the assumption of $\beta_i^+$ and $\beta_i^-$ being independent, the prior unconditional variance of $\beta_i$ is 
\begin{equation}
    \begin{split}
        \text{Var}(\beta_i) = \text{Var}(\text{max}(0, \beta_i^+)) + \text{Var}(\text{max}(0, \beta_i^-)) 
    \end{split}
\end{equation}
with 
\begin{equation}
    \begin{split}
        \text{Var}(\text{max}(0, \beta_i^+)) = \mathbb{E}(\text{max}(0, \beta_i^+)^2) - (\mathbb{E}(\text{max}(0, \beta_i^+)))^2.
    \end{split}
\end{equation}
The first term is calculated to be 
\begin{equation}
    \begin{split}
        \mathbb{E}(\text{max}(0, \beta_i^+)^2) &= \integral{0}{\infty} (\beta_i^+)^2 \mathcal{N}(\beta_i^+ \mid \mu, \rho^2) d\beta_i^+ \\
        &= \rho \mu \phi(\mu / \rho) + (\mu^2 + \rho^2) \Phi(\mu / \rho)
    \end{split}
\end{equation}
using integration by parts and results of Gaussian integrals. Let $\mu_2^* := \mathbb{E}(\text{max}(0, \beta_i^+)^2) = \mathbb{E}(\text{max}(0, \beta_i^-)^2) = \rho \mu \phi(\mu / \rho) + (\mu^2 + \rho^2) \Phi(\mu / \rho)$. Then, it follows that
\begin{equation}
    \text{Var}(\text{max}(0, \beta_i^+)) = \mu_2^* - (\mu_1^*)^2.
\end{equation}
Hence, if $(\rho^*)^2 := \text{Var}(\text{max}(0, \beta_i^+)) = \text{Var}(\text{max}(0, \beta_i^-)) = \mu_2^* - (\mu_1^*)^2$, the prior unconditional variance of $\beta_i$ can be written as 
\begin{equation}
    \text{Var}(\beta_i) = 2 (\rho^*)^2. 
\end{equation}

Finally, in order to calculate the prior conditional variance of $\beta_i$, i.e. $\text{Var}(\beta_i \mid \beta_i \neq 0)$, notice again that $\beta_i \neq 0$ whenever
\begin{itemize}
    \item $\beta_i^+ > 0$ and $\beta_i^- > 0$ with probability $P(\beta_i^+ > 0) P(\beta_i^- > 0) = (1 - \Phi(-\mu / \rho))^2$,
    \item $\beta_i^+ > 0$ and $\beta_i^- < 0$ with probability $P(\beta_i^+ > 0) P(\beta_i^- < 0) = (1 - \Phi(-\mu / \rho))\Phi(-\mu / \rho)$,
    \item $\beta_i^+ < 0$ and $\beta_i^- > 0$ with probability $P(\beta_i^+ < 0) P(\beta_i^- > 0) = \Phi(-\mu / \rho)(1 - \Phi(-\mu / \rho))$.
\end{itemize}
Using the definition of conditional probability, one can construct the density of $\beta_i^+ \mid \beta_i \neq 0$ (and similarly of $\beta_i^- \mid \beta_i \neq 0$) as $\mathcal{N}(\beta_i^+ \mid \mu, \rho^2) / P(\beta_i \neq 0)$. Based on the above, it can be seen that 
\begin{equation}
\begin{split}
    P(\beta_i \neq 0) &= 1 - P(\beta_i^+ < 0) P(\beta_i^- < 0) = 1- \Phi(-\mu / \rho)^2.
\end{split}
\end{equation}
Defining $c$ as $c = 1- \Phi(-\mu / \rho)^2 $, observe that 
\begin{equation}
    \begin{split}
        \mathbb{E}(\text{max}(0, \beta_i^+) \mid \beta_i \neq 0) &= \integral{0}{\infty} \beta_i^+ \frac{\mathcal{N}(\beta_i^+ \mid \mu, \rho^2)}{P(\beta_i \neq 0)} d\beta_i^+ \\
        &= \frac{1}{c}\mu^*_1
    \end{split}
\end{equation}
and 
\begin{equation}
    \begin{split}
        \mathbb{E}(\text{max}(0, \beta_i^+)^2 \mid \beta_i \neq 0) &= \integral{0}{\infty} (\beta_i^+)^2 \frac{\mathcal{N}(\beta_i^+ \mid \mu, \rho^2)}{P(\beta_i \neq 0)} d\beta_i^+ \\
        &= \frac{1}{c}\mu^*_2,
    \end{split}
\end{equation}
which leads to
\begin{equation}
    \text{Var}(\text{max}(0, \beta_i^+) \mid \beta_i \neq 0) = \frac{\mu^*_2}{c} - \frac{(\mu^*_1)^2}{c^2}. 
\end{equation}

Now, an important note is that when conditioning on $\beta_i \neq 0$, $\beta_i^+$ and $\beta_i^-$ are no longer independent due to the arguments from the bullet point list. For instance, given that $\beta_i \neq 0$, $\beta_i^-$ must be positive if $\beta_i^+$ is negative. Thus, 
\begin{equation}
\begin{split}
    \text{Var}(\beta_i \mid \beta_i \neq 0) &= \text{Var}(\text{max}(0, \beta_i^+) \mid \beta_i \neq 0) \\
    &+ \text{Var}(\text{max}(0, \beta_i^-) \mid \beta_i \neq 0) \\
    &- 2\text{ Cov}(\text{max}(0, \beta_i^+), \text{max}(0, \beta_i^-) \mid \beta_i \neq 0).
\end{split}
\end{equation}
In order to find the covariance term, one can argue with brute force by defining the joint density of $\beta_i^+$ and $\beta_i^-$ given $\beta_i \neq 0$. Also, the joint should be such that when looking at the marginal of e.g. $\beta_i^+$ given $\beta_i \neq 0$, the previous results hold (i.e. the marginal density is simply the unconditional prior density divided by $c$). Therefore, define the form of the joint density as
\begin{equation}
    f(\beta_i^+, \beta_i^- \mid \beta_i \neq 0) = f(\beta_i^+ \mid \beta_i \neq 0) f(\beta_i^- \mid \beta_i^+, \beta_i \neq 0).
\end{equation}
Since the marginal should have the properties and results from before, set 
\begin{equation}
    f(\beta_i^+ \mid \beta_i \neq 0) = \mathcal{N}(\beta_i^+ \mid \mu, \rho^2) / c.
\end{equation}
If $\beta_i^+ > 0$, $\beta_i^+$ can be anything on the real line for $\beta_i \neq 0$ to hold. In this case,
\begin{equation}
    f(\beta_i^- \mid \beta_i^+, \beta_i \neq 0) = \mathcal{N}(\beta_i^+ \mid \mu, \rho^2).
\end{equation}
If $\beta_i^+ < 0$, $\beta_i^-$ must be positive for $\beta_i \neq 0$ to hold. For this scenario, 
\begin{equation}
    f(\beta_i^- \mid \beta_i^+, \beta_i \neq 0) = \mathcal{N}(\beta_i^+ \mid \mu, \rho^2) \mathbb I (\beta_i^- > 0).
\end{equation}
The joint density can therefore be expressed as 
\begin{equation}
    f(\beta_i^+, \beta_i^- \mid \beta_i \neq 0) = \begin{cases}
        \frac{\mathcal{N}(\beta_i^+ \mid \mu, \rho^2)\mathcal{N}(\beta_i^- \mid \mu, \rho^2)}{c}, \text{ if }   \beta_i^+ > 0 \\
        \frac{\mathcal{N}(\beta_i^+ \mid \mu, \rho^2)\mathcal{N}(\beta_i^- \mid \mu, \rho^2) \mathbb I (\beta_i^- > 0)}{c}, \text{ if }   \beta_i^+ < 0.
    \end{cases}
\end{equation}
Using this result, $\mathbb{E}(\text{max}(0, \beta_i^+), \text{max}(0, \beta_i^-) \mid \beta_i \neq 0)$ can be calculated as follows:
\begin{equation}
    \begin{split}
        \mathbb{E}(\text{max}(0, \beta_i^+) \text{max}(0, \beta_i^-) \mid \beta_i \neq 0) &= \integral{-\infty}{\infty} \integral{-\infty}{\infty}  \text{max}(0, \beta_i^+) \, \text{max}(0, \beta_i^-) f(\beta_i^-, \beta_i^+ \mid \beta_i \neq 0) d\beta_i^+ d\beta_i^- \\
        &= \integral{0}{\infty} \integral{0}{\infty}  \beta_i^+ \beta_i^- f(\beta_i^-, \beta_i^+ \mid \beta_i \neq 0) d\beta_i^+ d\beta_i^- \\
        &= \frac{1}{c} \integral{0}{\infty} \beta_i^+ \mathcal{N}(\beta_i^+ \mid \mu, \rho^2) d\beta_i^+ \integral{0}{\infty} \beta_i^- \mathcal{N}(\beta_i^- \mid \mu, \rho^2) d\beta_i^- \\
        &= \frac{(\mu^*_1)^2}{c}.
    \end{split}
\end{equation}
Accordingly, the prior conditional covariance is \begin{equation}
\begin{split}
        \text{Cov}(\text{max}(0, \beta_i^+), \text{max}(0, \beta_i^-) \mid \beta_i \neq 0) &= \mathbb{E}(\text{max}(0, \beta_i^+) \text{max}(0, \beta_i^-) \mid \beta_i \neq 0) \\
        &- \mathbb{E}(\text{max}(0, \beta_i^+) \mid \beta_i \neq 0)E(\text{max}(0, \beta_i^-) \mid \beta_i \neq 0) \\
        &= \frac{(\mu^*_1)^2}{c} - \frac{(\mu^*_1)^2}{c^2}.
\end{split}    
\end{equation}
Finally, one arrives at the following result of the prior conditional variance:
\begin{equation} \label{eq:cond_var_beta_reg_lin_reg}
    \begin{split}
            \text{Var}(\beta_i \mid \beta_i \neq 0) &= 2 \left (\frac{\mu^*_2}{c} - \frac{(\mu^*_1)^2}{c^2} \right ) - 2 \left ( \frac{(\mu^*_1)^2}{c} - \frac{(\mu^*_1)^2}{c^2} \right ) \\
            &= \frac{2 (\mu_2^* - (\mu_1^*)^2)}{c} \\
            &= \frac{2 (\rho^*)^2}{c} =\frac{ \text{Var}(\beta_i)}{c}. 
    \end{split}
\end{equation}
Consequently, for a desired value of prior $P(\beta_i=0) = p^*$ and $\text{Var}(\beta_i \mid \beta_i \neq 0) = (\sigma^*)^2$, one can solve this system of two equations with respect to $\mu$ and $\rho$ numerically to obtain the prior hyperparameters required for achieving the anticipated behaviour of the prior. 

\subsection{Transformation to coefficients in original scale}
In Section \ref{sec:reg_lin_reg}, it is mentioned that the regression model is fitted on both scaled and centered response and covariates. The following results are standard, but they will be presented here for completeness. Let 
\begin{align}
    \mathbf{x}_i^s & := \frac{\mathbf{x}_i-\Bar{\mathbf{x}}_i}{s_{\mathbf{x}_i}}, \,  i = 1 \dots p, \\
    \mathbf{y}^s &:= \frac{\mathbf{y}-\Bar{\mathbf y}}{s_{\mathbf y}},
\end{align}
where $\Bar{\mathbf x}_i$ is the sample mean of the $i$-th covariate observed in the data set and $s_{\mathbf{x}_i}$ the sample standard deviation of the corresponding covariate (similar definitions for $\mathbf{y}$). Using the scaled covariates and response, the resulting linear model will not contain an intercept term. Hence, for a new observation with a  scaled covariate vector $(\Tilde{x}_1^s, \dots, \Tilde{x}_p^s)$, 
\begin{equation}
    \Tilde{y}^s = \sum_{i=1}^p \beta_i^s \Tilde{x}_i^s,
\end{equation}
which after inserting the two prior defined relations yields 
\begin{equation}
\frac{\Tilde{y}-\Bar{\mathbf y}}{s_{\mathbf y}} = \sum_{i=1}^p \beta_i^s \frac{\Tilde{x}_i-\Bar{\mathbf{x}}_i}{s_{\mathbf{x}_i}}. 
\end{equation}
Then, simple manipulations give 
\begin{equation}
    \Tilde{y} = \Bar{\mathbf y} - s_{\mathbf y} \sum_{i=1}^p \frac{\beta_i^s \Bar{\mathbf{x}}_i}{s_{\mathbf{x}_i}} + \sum_{i=1}^p s_{\mathbf y} \frac{\beta_i^s}{s_{\mathbf{x}_i}} \Tilde{x}_i,
\end{equation}
thus implying that the intercept and coefficient parameters in the original scale are as follows: 
\begin{align}
    \beta_0 &= \Bar{\mathbf y} - s_{\mathbf y} \sum_{i=1}^p \frac{\beta_i^s \Bar{\mathbf{x}}_i}{s_{\mathbf{x}_i}}, \\
    \beta_i &= s_{\mathbf y} \frac{\beta_i^s}{s_{\mathbf{x}_i}}.
\end{align}

\end{document}